\documentclass[aps,prx,twocolumn,showpacs,superscriptaddress,floatfix]{revtex4-2}  % for review and submission
\usepackage{ucs}
\usepackage{graphicx}
\usepackage{subcaption}  % for subfigures
\usepackage{amsmath}
\usepackage{natbib}
\usepackage[normalem]{ulem}
\usepackage{graphicx}  % needed for figures
\usepackage{dcolumn}   % needed for some tables
\usepackage{tikz}
\usetikzlibrary{positioning}
\usepackage{bbold}
\usepackage{booktabs} 
\usepackage{bm}        % for math
\usepackage{amssymb}   % for math
\usepackage{amsmath}
\usepackage{color}
\usepackage{times}
\usepackage{verbatim}
\usepackage{mathrsfs}
\usepackage{hyperref}
\usepackage{ulem}
\hypersetup{colorlinks = True, urlcolor=blue, linkcolor=blue, citecolor=blue}
\newcommand{\bg}{\begin{pmatrix}}
\newcommand{\ed}{\end{pmatrix}}

\usepackage{cancel}
\usepackage{comment}

\begin{document}

\title{Band geometry, Anderson's theorem, and vortex nucleation in unconventional 2D superconductors}
\author{Vladislav Poliakov}\email{vlad\_p@mit.edu}
\affiliation{Department of Physics, Massachusetts Institute of Technology, Cambridge, MA 02139, USA}
\author{Inti Sodemann Villadiego}
\affiliation{Institut für Theoretische Physik, Universität Leipzig, Brüderstraße 16, 04103 Leipzig, Germany}
\author{T. Senthil}
\affiliation{Department of Physics, Massachusetts Institute of Technology, Cambridge, MA 02139, USA}

\begin{abstract}
    Motivated by the recent discovery of chiral superconductors in $2d$ materials, we study the effects of disorder on superconductivity in materials with non-trivial band geometry. First, we show how Anderson's theorem and the conventional Abrikosov-Gorkov description of disorder-induced pair breaking are modified by the structure of the normal-state Bloch functions. We apply these results to twisted MoTe$_2$ and  show that a  BCS pairing mechanism for the observed superconductor  is in tension with the measured normal state residual resistivity. Second, we study whether random potential disorder can nucleate superconducting vortices in a chiral time-reversal-breaking superconductor. We show that, for a superconductor stabilized by the usual BCS mechanism, weak disorder does not induce a vortex glass phase, in contrast to superconductivity formed from an anyon fluid.
\end{abstract}

\maketitle

\section{Introduction} 
The discovery of superconductivity in 2d materials, particularly in moir\'e systems, has led to a number of questions on the mechanism stabilizing these states, and their phenomenology in experiments \cite{Andrei_the_2021, Andrei_graphene_2020}. It has increasingly become clear that several of these superconductors have unconventional pairing symmetries, {\it i.e.}, have order parameters that are not spin-singlet s-wave \cite{Oh_evidence_2021, Kim_evidence_2022, Park_robust_2022, Lake_pairing_2022, Biswas_andreev_2025, Park_experimental_2026, Sun_nodal_2026, Biswas_probing_2026,graphene_sc,Xu2025_SCdopeTMD,Hua_multi-knob_2026}. Of particular interest to us is 
the discovery of chiral superconductivity in N-layer rhombohedral graphene (RNG) with $N\geq 4$ \cite{graphene_sc,Hua_multi-knob_2026} and in twisted MoTe$_2$ \cite{Xu2025_SCdopeTMD}.  In the context of this paper (and in the experimental papers mentioned above), chiral superconductivity means that the state breaks time-reversal symmetry in the orbital motion of the electrons. In these two-dimensional ($2d$) materials, this is achieved by spontaneous polarization of the valley degree of freedom, as evidenced both in RNG and in twisted MoTe$_2$.

At the outset it is important to recognize that RNG and tMoTe$_2$ are fairly different systems. In the former, there is no moir\'e potential, and the `normal' state above $T_c$ has a Fermi surface seen in quantum oscillations (though details of its shape are not settled) \cite{graphene_sc, Kalantre_fermiology_2026}. RNG is also very clean and the residual resistivity of the normal state is small compared to the resistance quantum $h/e^2$. In contrast, the superconductivity seen in tMoTe$_2$ occurs in a Chern band produced by a moir\'e potential, and at a band filling $\nu = 2/3 + \delta$ (with $\delta$ small), right next to the Fractional Quantum Anomalous Hall (FQAH) state that exists in a range of twist angles. Further, when compared with RNG,  tMoTe$_2$ is relatively disordered: the residual resistivity of the normal state is a sizeable fraction of $h/e^2$. Assuming a standard Drude metal, the corresponding mean free path $l$ is smaller than the reported superconducting coherence length $\xi$ placing tMoTe$_2$ in the dirty limit.  Indeed, a comparison with other superconductors in 2d materials (see Appendix \ref{app: overview}) reveals that, among those expected to have unconventional pairing symmetry, tMoTe$_2$ has a relatively large residual resistivity in its normal state, and the smallest value of $l/\xi$ (assuming a standard normal state with Drude transport).

The discovery of the FQAH state \cite{Cai2023_FQAHTMD,Park2023_FQAH_TMD,Xu2023_FQAHTMD,Zeng2023_FQAHTMD,Lu2023_FQAHPenta,Lu2025_EQAH} in tMoTe$_2$ and in RNG aligned with hBN led to a revival and refinement~\cite{Shi2024_doping,Kim2024_anyonSC,Shi2025_dopeMR, Zhang2025_SU(3)1dope, Shi2025_anyon_delocalization,  Nosov2025_plateau,Pichler2025_anyonSC,Han2025_anyonexciton,Kuhlenkamp2025_HofHubb,Shi2025_nonAbelian_TSC,lotrivc2026phases,Fan2026_weakpairing_SC,Wang2026_U3anyonSC,shi2026superconductivity} of the old notion of ``anyon superconductivity"~\cite{Laughlin1988_anyonSC,Lee1989_anyonSC,Fetter1989_anyonSC_RPA,Chen1989_anyonSC,Wen1990_anyonSC,Tang2013_anyonSC}. Following the discovery of superconductivity in tMoTe$_2$, some of this literature \cite{Shi2025_anyon_delocalization,Nosov2025_plateau,Shi2025_nonAbelian_TSC,Fan2026_weakpairing_SC,shi2026superconductivity,senthil2026fractionalized,han2026single} explored explanations based on the physics of an anyon fluid in this system. Invoking the anyon fluid requires that the carrier density in the superconducting region be ``small", {\it i.e.}, be set by the deviation $\delta$ of the filling from $2/3$. An alternative framework for an explanation postulates \cite{xu2025chiral,Guerci_from_2025,guerci2026topological} that as the chemical potential is tuned away from filling $\nu = 2/3$ toward $\nu = 1$, there is a strong first-order transition to a hole metal with carrier density set by $1-\nu$. This `large' Fermi surface metal might then undergo a pairing instability to form a superconductor. Arguments have been advanced \cite{Shi2025_anyon_delocalization,Nosov2025_plateau,Guerci_from_2025,Fan2026_weakpairing_SC,senthil2026fractionalized} in favor of both of these possibilities which we will not dwell into here \footnote{We note that other recent works that either demonstrate superconductivity near the FQAH through numerical study of tMoTe$_2$ models \cite{guerci2026topological,Wang_repulsion-driven_2026}, or analyze \cite{lotrivc2026fractional} experimental data on tMoTe$_2$ do not really weigh in on this question definitively}. Rather our focus will be on disorder effects in the `large' Fermi surface scenario which need to be addressed carefully given the large residual resistivity.

For conventional systems, there is an Anderson theorem which (within BCS theory) shows that ordinary $s$-wave superconductivity is unaffected by disorder while higher angular momentum paired states are strongly sensitive to it \cite{Anderson1959}. One of our goals, motivated by the experiments mentioned above, is to understand how this is modified, if at all, by the presence of non-trivial band geometry. More generally, when the bands in the normal state have a non-trivial structure of Bloch functions, of which the Berry curvature and quantum metric are a manifestation, how is Anderson's theorem modified? A similar question was considered in Ref. ~\cite{Sedov_quantum_2025} which analyzed the dependence of the superconducting temperature on the disorder strength for arbitrary band dispersions, including those asymmetric under momentum inversion: $\xi_{\boldsymbol{p}}\neq \xi_{-\boldsymbol{p}}$. These authors obtained an analytical result using a perturbative expansion in the impurity scattering rate $1/\tau$ and analyzed finite disorder strengths numerically. In the present work, however, we focus on a simpler rotationally symmetric dispersion (approximately true for tMoTe$_2$) , which allows us to obtain an explicit Abrikosov-Gor'kov-type expression for disorder strength comparable to superconducting energy scale. Our results give an intuitive understanding of $T_c$ suppression and the separate roles of the quantum metric and Berry curvature for simple models. Furthermore, we relate the scattering rate to the residual resistivity and derive a criterion for the critical residual resistivity above which the unconventional superconductivity is completely suppressed. Applying these results to tMoTe$_2$, we argue that the high residual resistivity in that system is in tension with a conventional BCS explanation of the superconductivity, supporting the consideration of alternative non-BCS  mechanisms.

Another question relates to the role of orbital magnetization which may be present when the state breaks time reversal symmetry. This is definitely true in the chiral superconductors in RNG and tMoTe$_2$. Within standard Ginzburg-Landau theory, the orbital magnetization can couple linearly to  a random potential and to the vorticity of the superconductor. This raises the possibility that such a random potential could spontaneously nucleate vortices in the superconducting state. How then does a chiral superconductor react to a weakly random potential? We address this by mapping the problem to well-studied random $2d$ classical $XY$ models \cite{Rubinstein_two-dimensionalmmlmath_1983,nattermann1995absence,scheidl1997glassy,carpentier1998disordered}. We show that for a superconductor stabilized by the usual BCS mechanism, the appropriate model is a weakly random phase shift $XY$ model \cite{Korshunov_2006}. In this model, the superconducting state is stable at weak randomness and there is a sharp finite-$T$ phase transition. In contrast, for superconductors typically stabilized through the formation of an anyonic fluid, the appropriate model is a random flux $2d$ XY model where the SC only survives as an `Anomalous Vortex Glass' (AVG) \cite{Shi2025_anyon_delocalization, senthil2026fractionalized}, and has no finite-$T$ phase transition. 
\par
This paper is structured as follows. In section II we derive the dependence of the superconducting critical temperature on disorder strength and illustrate how band geometry and Berry curvature modify the disorder-induced pair breaking, using form factors motivated by the  Lowest Landau Level and rhombohedral graphene as examples. In section III we recast these results in terms of the residual resistivity and find an upper bound on it compatible with superconductivity. In section IV we apply this framework to twisted MoTe$_2$. In section V we go beyond mean-field theory and study the effect of disorder on chiral superconductivity with orbital magnetization. We show that, at zero temperature, weak disorder produces only an exponentially small density of vortex-antivortex pairs, while the ground state retains long-range superconducting order. In section VI we contrast the response to disorder of a BCS superconductor with that of anyon-induced superconductors. Finally, in Appendix \ref{app: overview} we survey properties of several superconductors in $2d$ materials which emphasizes the relative robustness of the tMoTe$_2$ superconductor despite the short mean free path, assuming a standard normal state with Drude transport fo electronic quasiparticles.

\section{Effects of disorder in BCS mean field}
We begin with the standard BdG Hamiltonian for electrons in a single band. We suppress the spin and other internal indices and work within a single pairing channel. The discussion below applies both to even- and odd-parity pairing channels. The Hamiltonian can be written as:
\begin{eqnarray}\label{initial bdg}
\hat H_0
=
\frac{1}{2}
\int
\Psi_{\boldsymbol{p}}^\dagger
\begin{pmatrix}
\xi_{\boldsymbol{p}} & \Delta_{\boldsymbol{p}} \\
\Delta^*_{\boldsymbol{p}} & -\xi_{\boldsymbol{p}}
\end{pmatrix}
\Psi_{\boldsymbol{p}} \frac{d^2 p}{(2\pi)^2}. 
\end{eqnarray}
Here $\Psi_{\boldsymbol{p}}^\dagger =
\begin{pmatrix}
\psi_{\boldsymbol{p}}^\dagger & \psi_{-\boldsymbol{p}}
\end{pmatrix}$ is a Nambu vector defined in terms of Bloch operators in a band of interest: 
\begin{eqnarray}
    \psi_{\boldsymbol{p}}^\dagger=\int e^{i\boldsymbol{p}\boldsymbol{r}}u_{\boldsymbol{p}}(\boldsymbol{r})c_{\boldsymbol{r}}^\dagger d^2r.
\end{eqnarray}
Here $c_{\boldsymbol{r}}$ is the real space annihilation operator and $u_{\boldsymbol{p}}(\boldsymbol{r})$ is a periodic part of the Bloch function. In principle, each function $u_{\boldsymbol{p}}(\boldsymbol{r})$ can be multiplied by a certain phase factor $e^{i\phi_{\boldsymbol{p}}}$, after which the dispersion term remains invariant, but $\Delta_{\boldsymbol{p}}$ acquires a phase. That means that the Hamiltonian is invariant under the following gauge transformation:
\begin{eqnarray}\label{gauge transformation}
    u_{\boldsymbol{p}}(\boldsymbol{r}) \rightarrow u_{\boldsymbol{p}}(\boldsymbol{r}) e^{i\phi_{\boldsymbol{p}}}, \notag \\
    \Delta_{\boldsymbol{p}} \rightarrow \Delta_{\boldsymbol{p}} e^{-i\phi_{\boldsymbol{p}}-i\phi_{\boldsymbol{-p}}}.
\end{eqnarray}
We fix a choice of gauge for all further calculations. That will affect the structure of $\Delta_{\boldsymbol{p}}$ in momentum space, but all the physical quantities such as critical temperature and band gaps will remain invariant.

Let us introduce a random disorder potential  
\begin{eqnarray}
    \hat H & = & \hat H_0 + \hat V, \notag \\
    \hat V& = & \frac{1}{2}
\int 
\begin{pmatrix}
c_{\boldsymbol{r}}^\dagger & c_{\boldsymbol{r}}
\end{pmatrix}
\begin{pmatrix}
V_{\boldsymbol{r}} & 0 \\
0 & -V_{\boldsymbol{r}}
\end{pmatrix}
\begin{pmatrix}
c_{\boldsymbol{r}} \\
c^\dagger_{\boldsymbol{r}}
\end{pmatrix}
d^2 r,
\end{eqnarray}
where
 $V_{\boldsymbol{r}}$ is a Gaussian distributed random potential with zero mean and correlation function given by 
\begin{eqnarray}
    \langle V_{\boldsymbol{r}_1} V_{\boldsymbol{r}_2} \rangle=g^2 \delta (\boldsymbol{r}_1-\boldsymbol{r}_2).
\end{eqnarray}
After projecting into a band of interest, the disorder potential acquires two form factors dependent on band geometry: 
\begin{equation}
\hat V
=
\frac{1}{2}
\int 
\Psi_{\boldsymbol{k}}^\dagger
V_{\boldsymbol{k}-\boldsymbol{p}}
\left(
\begin{smallmatrix}
\Lambda_{\boldsymbol{k},\boldsymbol{p}} & 0\\
0 & -\Lambda_{-\boldsymbol{p},-\boldsymbol{k}}
\end{smallmatrix}
\right)
\Psi_{\boldsymbol{p}}
\frac{d^2 p}{(2\pi)^2}\frac{d^2 k}{(2\pi)^2},
\end{equation}
Here $\Lambda_{\boldsymbol{k},\boldsymbol{p}}=\langle u_{\boldsymbol{k}}|u_{\boldsymbol{p}}\rangle$ is the standard expression for the overlap of two Bloch functions.
 We want to address how a spatially homogeneous mean-field order parameter is affected by the presence of impurity potential. To that end, we follow the approach introduced in \cite{Vojta2005} and include disorder-averaged self-energy corrections to the Nambu  Green function:
\begin{eqnarray}
    G(i\omega_n,\boldsymbol{p})=\frac{1}{i\omega_n-H_{\text{BdG}}(\boldsymbol{p})-\Sigma(i \omega_n,\boldsymbol{p})}.
\end{eqnarray}
Here $H_{\text{BdG}}$ is a $2\times2$ matrix from eqn \eqref{initial bdg}.

\begin{figure}[t]
    \centering
    \begin{tikzpicture}[
    baseline=(current bounding box.center),
    bare/.style={line width=0.6pt},
    full/.style={line width=2.0pt},
    impurity/.style={dashed, line width=1.0pt},
    selfenergy/.style={
        circle,
        draw,
        line width=0.9pt,
        minimum size=9mm,
        inner sep=0pt,
        font=\Large
    }
]

% =========================================================
% Dyson equation: G = G0 + G0 Sigma G
% =========================================================

% Full G
\draw[full] (0,0) -- (1.3,0);

\node at (1.85,0) {$=$};

% Bare G0
\draw[bare] (2.35,0) -- (3.65,0);

\node at (4.15,0) {$+$};

% G0 Sigma G
\draw[bare] (4.65,0) -- (5.55,0);

\node[selfenergy] (Sigma1) at (6.0,0) {$\Sigma$};

% touches Sigma circle
\draw[full] (6.45,0) -- (7.75,0);

% =========================================================
% Self-consistent Born equation for Sigma
% =========================================================

\node[selfenergy] (Sigma2) at (0.55,-2.0) {$\Sigma$};

\node at (1.4,-2.0) {$=$};

% longer self-energy propagator
\draw[bare] (2.0,-2.0) -- (2.55,-2.0);
\draw[full] (2.55,-2.0) -- (5.35,-2.0);
\draw[bare] (5.35,-2.0) -- (5.9,-2.0);

% impurity correlator
\draw[impurity]
    (2.55,-2.0)
    .. controls (3.1,-0.85) and (4.8,-0.85) ..
    (5.35,-2.0);

% impurity vertices
\fill (2.55,-2.0) circle (1.4pt);
\fill (5.35,-2.0) circle (1.4pt);

\end{tikzpicture}
    \caption{
        A self-consistent Green's function calculation. Thin and thick lines correspond to the bare and fully dressed superconducting Green's function respectively, and the dashed line represents disorder-averaged scattering on impurities. After ignoring the diagrams with crossed impurity lines, the fully dressed Green's function can be found by using Dyson's equation (top line) and self-consistent self-energy (bottom line).
    }
    \label{fig:Dyson}
\end{figure}
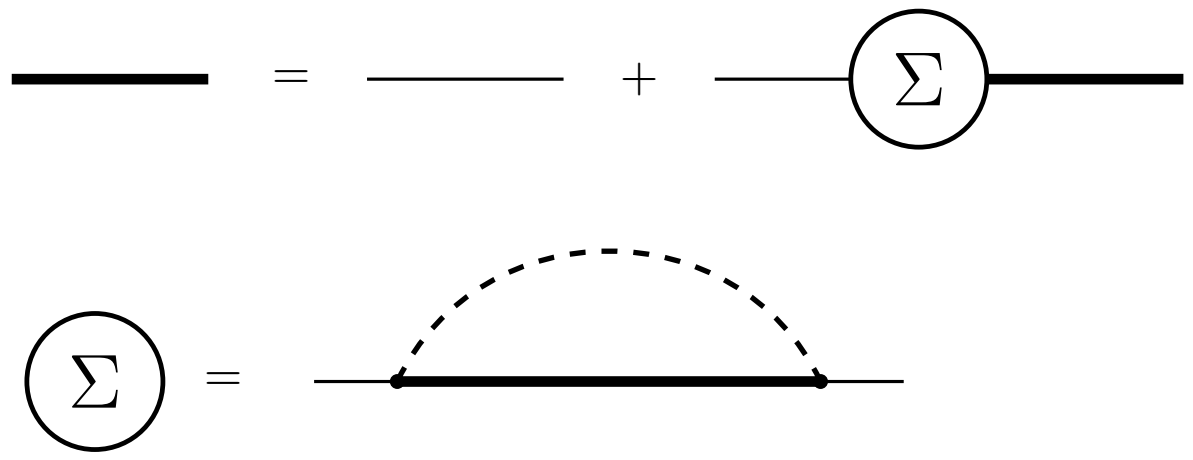

In this problem we have three different energy scales: the Fermi energy $E_F$, the disorder scattering rate $\frac{1}{\tau}=2\pi g^2 \nu_0$, where $\nu_0$ is the density of states at the Fermi level and the superconducting critical temperature in the clean limit $T_0$. We assume that the Fermi energy is the largest parameter, but the scattering rate and critical temperature may be comparable:
\begin{eqnarray}
    E_F \gg T_0 \sim \frac{1}{\tau}.
\end{eqnarray} 
In order to study the mean-field superconducting transition we need to write a gap equation in terms of disorder-renormalized Green functions.
Using standard disorder-averaged diagrammatic techniques, one can see that diagrams with the crossed dashed lines contain a small factor $\frac{1}{\tau E_F}$ or $\frac{T_0}{E_F}$, so they can be neglected. The summation of all the other diagrams may be incorporated in a self-consistent calculation shown in Fig.~\ref{fig:Dyson}. The first line represents a standard Dyson equation:
$G^{-1}(i \omega_n, \boldsymbol{p})=G_0^{-1}(i \omega_n, \boldsymbol{p})-\Sigma(i \omega_n, \boldsymbol{p})$ while the second line finds the self-energy self-consistently with the dressed Green's function. For a BdG superconductor defined by the Hamiltonian \eqref{initial bdg} the Green's function takes the form
\begin{eqnarray}
    G(i\omega_n,\boldsymbol{p})=-\frac{i\tilde{\omega}_n \tau_0
    + \tilde{\Delta}'_{\boldsymbol{p}} \tau_x
    + \tilde{\Delta}''_{\boldsymbol{p}} \tau_y
    + \tilde{\xi}_{\boldsymbol{p}} \tau_z}
   {\tilde{\omega}_n^2 + \tilde{\xi}_{\boldsymbol{p}}^2 + |\tilde{\Delta}_{\boldsymbol{p}}|^2},
\end{eqnarray}
where 
\begin{eqnarray}\label{shifted variables}
    i\tilde{\omega}_n=i\omega_n-\Sigma_0(i\omega_n,\boldsymbol{p}), \notag \\    \tilde{\Delta}'_{\boldsymbol{p}}=\Delta'_{\boldsymbol{p}}+\Sigma_x(i\omega_n,\boldsymbol{p}), \notag \\    \tilde{\Delta}''_{\boldsymbol{p}}=\Delta''_{\boldsymbol{p}}+\Sigma_y(i\omega_n,\boldsymbol{p}), \notag \\
\tilde{\xi}_{\boldsymbol{p}}=\xi_{\boldsymbol{p}}+\Sigma_z(i\omega_n,\boldsymbol{p}), 
\end{eqnarray}
and $\Sigma_i$ is the $i$-th component of the self-energy in an expansion over Nambu-space Pauli matrices: $\Sigma=\tau_0 \Sigma_0+\tau_x \Sigma_x+\tau_y \Sigma_y+\tau_z \Sigma_z$. The self-energy then can be written as 
\begin{widetext} 
\begin{equation}\label{self energy}
\Sigma(i\omega_n,\boldsymbol{p}) 
=  -\int \frac{g^2}{\tilde{\omega}_n^2 + \tilde{\xi}_{\boldsymbol{k}}^2 + |\tilde{\Delta}_{\boldsymbol{k}}|^2} \,
\begin{pmatrix}
\Lambda_{\boldsymbol{p},\boldsymbol{k}} & 0 \\[1mm]
0 & -\Lambda_{-\boldsymbol{k},-\boldsymbol{p}}
\end{pmatrix}
\begin{pmatrix}
i\tilde{\omega}_n + \tilde{\xi}_{\boldsymbol{k}} & \tilde{\Delta}_{\boldsymbol{k}} \\[1mm]
\tilde{\Delta}_{\boldsymbol{k}}^* & i\tilde{\omega}_n - \tilde{\xi}_{\boldsymbol{k}}
\end{pmatrix}
\begin{pmatrix}
\Lambda_{\boldsymbol{k},\boldsymbol{p}} & 0 \\[1mm]
0 & -\Lambda_{-\boldsymbol{p},-\boldsymbol{k}}
\end{pmatrix} \frac{d^2k}{(2\pi)^2}.
\end{equation}
\end{widetext}
From now on let us focus on a circular Fermi surface. Near the superconducting transition point we retain terms only to linear order in $\Delta_{\boldsymbol{k}}$. Within the usual weak-coupling approximation we restrict our analysis to a narrow shell around the Fermi surface and neglect the radial dependence of density of states and band geometry form factors. Assuming particle-hole symmetry within the shell, the terms linear in $\xi_{\boldsymbol{k}}$ vanish upon radial integration, so that $\Sigma_z=0$ up to a constant shift in chemical potential. For the same reason $\Sigma_0(i\omega_n,\boldsymbol{p})$ does not depend on the radial component of momentum. By performing the radial integral in Eq. \eqref{self energy} and substituting it into Eq. \eqref{shifted variables} we arrive at the following self-consistency equations:
\begin{align}\label{new omega and delta}
    &\tilde{\omega}_n=\omega_n+\frac{1}{2\tau}\frac{\tilde{\omega}_n}{|\tilde{\omega}_n|}\int
    \frac{
|\Lambda_{\boldsymbol{p},\boldsymbol{k}}|^2
    +
    |\Lambda_{-\boldsymbol{p},-\boldsymbol{k}}|^2
    }{2}
    \frac{d\theta_{\boldsymbol{k}}}{2\pi}, \notag \\
    &\tilde{\Delta}_{\boldsymbol{p}}=\Delta_{\boldsymbol{p}}+\frac{1}{2\tau}\frac{1}{|\tilde{\omega}_n|}
    \int
\Lambda_{\boldsymbol{p},\boldsymbol{k}}
    \Lambda_{-\boldsymbol{p},-\boldsymbol{k}}
    \tilde{\Delta}_{\boldsymbol{k}}
    \frac{d\theta_{\boldsymbol{k}}}{2\pi}.
\end{align}
For arbitrary band geometry, both $\tilde{\omega}_n$ and $\tilde{\Delta}_{\boldsymbol{p}}$ acquire nontrivial angular dependence determined by disorder strength and band geometry form factors $\Lambda_{\boldsymbol{p},\boldsymbol{k}}$. Let us now assume that, like the Fermi surface itself, the band geometry is also rotationally symmetric, so that $\tilde{\omega}_n$ is independent of the direction of $\boldsymbol{p}$. The angular dependence of $\tilde{\Delta}_{\boldsymbol{p}}$ is non-trivial, since its phase depends on the direction of $\boldsymbol{p}$ and is also affected by the gauge choice introduced in Eq.~\eqref{gauge transformation}. In order to account for this, let us separate the gauge-invariant magnitude of the order parameter from its gauge-dependent angular structure: $\Delta_{\boldsymbol{p}}\equiv\Delta g_{\boldsymbol{p}}$, where the angular part is normalized according to $\int g_{\boldsymbol{p}}^*g_{\boldsymbol{p}} \frac{d\theta_p}{2\pi}=1$. Then the second equation above can be written as $\tilde{\Delta}_{\boldsymbol{p}}=\Delta_{\boldsymbol{p}}(1-\frac{1}{2\tau \tilde{\omega}_n}\int g^*_{\boldsymbol{q}}
\Lambda_{\boldsymbol{q},\boldsymbol{k}}
    \Lambda_{-\boldsymbol{q},-\boldsymbol{k}}
    g_{\boldsymbol{k}}
    \frac{d\theta_{\boldsymbol{q}}}{2\pi}\frac{d\theta_{\boldsymbol{k}}}{2\pi} )^{-1}$. Here, the phase structure of $\tilde{\Delta}_{\boldsymbol{p}}$ is inherited from $\Delta_{\boldsymbol{p}}$, but the magnitude of the order parameter is gauge-invariant. Although the phase of $g_{\boldsymbol{p}}$ depends on the gauge choice, if the order parameter is fully gapped, the phase winding of $g_{\boldsymbol{p}}$ cannot be changed by a gauge transformation that is smooth and nonsingular throughout the entire Fermi sea. Formally, if we write $\Delta_p = |\Delta_p| e^{i \alpha_p}$, then under the gauge transformation in Eqn. \ref{gauge transformation}, $\alpha_p \rightarrow \alpha_p - \phi_p - \phi_{-p}$. Thus the integer phase winding $m$ defined through 
    \begin{equation} 
    \oint_{FS} d \alpha_p = 2\pi m 
    \end{equation} 
     is a well-defined gauge invariant quantity (the integral in the equation above is over the Fermi surface). Therefore, let us assume that in a particular gauge, we can choose $\Delta_{\boldsymbol{p}}=e^{im \theta_{\boldsymbol{p}}}\Delta_m$, in which case the self-consistency equation has the following solutions:

\begin{align}\label{renorm w and delta}
&\tilde{\omega}_n=\omega_n+\frac{\omega_n}{|\omega_n|}\frac{f_{\text{sym}}}{2\tau}, \notag \\
    &\tilde \Delta_m= \Delta_m \frac{|\omega_n|+\frac{f_{\text{sym}}}{2\tau}}{|\omega_n|+\frac{f_{\text{sym}}}{2\tau}-\frac{f_{m}}{2\tau}},
\end{align}
where all information about band geometry is encoded in two factors:
\begin{align}\label{fsym and fm definition}
    & f_{\text{sym}}=\int \frac{|\Lambda_{\boldsymbol{k},\boldsymbol{p}}|^2+|\Lambda_{-\boldsymbol{k},-\boldsymbol{p}}|^2}{2} \frac{d\theta_{\boldsymbol{p}}}{2\pi}\frac{d\theta_{\boldsymbol{k}}}{2\pi}, \notag \\
    & f_m=
    \int
    \Lambda_{\boldsymbol{k},\boldsymbol{p}}
    \Lambda_{-\boldsymbol{k},-\boldsymbol{p}}
    e^{im(\theta_{\boldsymbol{p}}-\theta_{\boldsymbol{k}})}
    \frac{d\theta_{\boldsymbol{p}}}{2\pi}
    \frac{d\theta_{\boldsymbol{k}}}{2\pi}.
\end{align}
Now, given an attractive interaction
\begin{eqnarray}\label{hint}
    \hat H_{\text{int}}=\frac{1}{2}\int U(\boldsymbol{k},\boldsymbol{p}) \psi_{\boldsymbol{p}}^\dagger\psi_{-\boldsymbol{p}}^\dagger \psi_{-\boldsymbol{k}}\psi_{\boldsymbol{k}} \frac{d^2k}{(2\pi)^2} \frac{d^2p}{(2\pi)^2},
\end{eqnarray}
the gap equation can be written as
\begin{eqnarray}\label{gapeqn}
    \Delta_{\boldsymbol{k}}=-\int U(\boldsymbol{k},\boldsymbol{p})T\sum_{\omega_n} \frac{\tilde{\Delta}_{\boldsymbol{p}}}{\tilde{\omega}_n^2 + \tilde{\xi}_{\boldsymbol{p}}^2 + |\tilde{\Delta}_{\boldsymbol{p}}|^2} \frac{d^2p}{(2\pi)^2} .
\end{eqnarray}
To obtain the critical temperature $T_c$, we linearize the above equation by neglecting the order parameter in the denominator, and do the radial integral to  get
\begin{eqnarray}
    \Delta_k=-\nu_0 \int U(\boldsymbol{k},\boldsymbol{p})\pi T_c \sum_{\omega_n}\frac{\tilde{\Delta}_{\boldsymbol{p}}}{|\tilde{\omega}_n|}.
\end{eqnarray}
By rewriting in angular momentum basis and substituting $\tilde{\omega}_n,\tilde{\Delta}_m$ from \eqref{renorm w and delta} we get 
\begin{eqnarray}
    -\frac{1}{\lambda_m \nu_0}=2\pi T_c \sum_{\omega_n>0}\frac{1}{\omega_n+\frac{f_{\text{sym}}}{2\tau}-\frac{f_{m}}{2\tau}},
\end{eqnarray}
where $\lambda_m$ is an angular eigenvalue of the attractive interaction $U(\boldsymbol{k}, \boldsymbol{p})$.
After calculating the sum over frequencies and comparing it to the clean limit  $\frac{1}{\tau}=0$ case we arrive at the following equation, which allows to express the dependence of the critical temperature on the impurity scattering time: 
 \begin{eqnarray}\label{crit temp eqn with disorder}
    \ln \left(\frac{T_0}{T_c}\right)= \psi\left(\frac{1}{2}+\frac{f_{\text{sym}}-f_m}{4\pi \tau T_c}\right)-\psi\left(\frac{1}{2}\right) ,
\end{eqnarray}

where $\psi(x)$ is the digamma function and $T_c$ ($T_0$) is the mean field critical temperature of the superconductor with (without) impurities. This equation is the central result of this section. For a trivial band with momentum-independent form factors, we would have $f_{\text{sym}}=1$ and $f_m=\delta_{m,0}$.  Then, for $m=0$, i.e. s-wave superconductivity,  the right hand side vanishes which means that $T_c$ is unaffected by disorder,  reproducing Anderson's theorem  \cite{Anderson1959}. For $m \neq 0$, we also reproduce the universal Abrikosov-Gorkov formula for pair breaking disorder \cite{Mineev1999}. Equation \eqref{crit temp eqn with disorder} predicts a total suppression of superconductivity when the impurity scattering rate exceeds a critical threshold 
\begin{eqnarray}
    \frac{1}{\tau_c}= \frac{\pi e^{-\gamma}}{f_{\text{sym}}-f_m} T_0,
\end{eqnarray}
where $\gamma\approx 0.577$ is the Euler constant. All model-dependent information is now encoded in two factors $f_{\text{sym}}$ and $f_m$ defined in terms of band form factor $\Lambda_{\boldsymbol{k},\boldsymbol{p}}=\langle u_{\boldsymbol{k}}| u_{\boldsymbol{p}} \rangle$ in equations \eqref{fsym and fm definition}. Nontrivial band geometry and Berry curvature modify $f_{\text{sym}}$ and $f_m$, and therefore the disorder sensitivity of different pairing channels.
\par
 In the remainder of this section we will develop  intuition for how different band geometries affect these parameters and thus how superconductivity is suppressed by disorder. We will start by considering the case where the form factor is similar to that in the  Lowest Landau Level (LLL): this yields  a model of an ideal Chern band with uniformly distributed Berry curvature. Then we will consider the toy $2$-band model of n-layer rhombohedral graphene with a small displacement field, which has a concentrated  distribution of Berry curvature. Finally, we will apply these results to the tMoTe$_2$ \cite{Xu2025_SCdopeTMD} experiment and discuss the relation between the superconducting critical temperature and residual resistance in the normal state.

\subsection{Uniformly spread Berry curvature: lowest Landau level form factor}
Equation ~\eqref{crit temp eqn with disorder} allows us to estimate the mean field critical temperature knowing the disorder scattering time and the band form factors.
 
\begin{figure*}[t]
    \centering

    %================ Top row =================%

    % Panel (a)
    \begin{subfigure}[b]{0.48\textwidth}
        \centering
        \includegraphics[width=\linewidth]{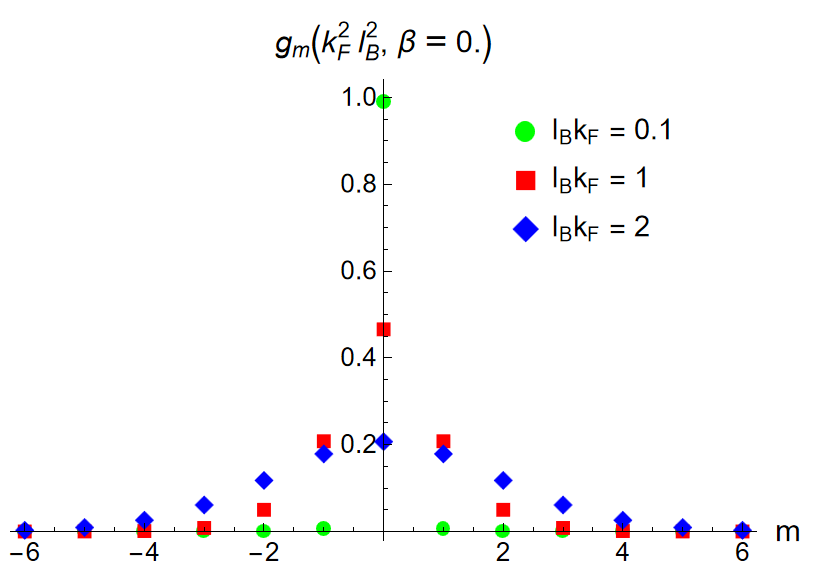}
    \end{subfigure}
    \hfill
    % Panel (b)
    \begin{subfigure}[b]{0.48\textwidth}
        \centering
        \includegraphics[width=\linewidth]{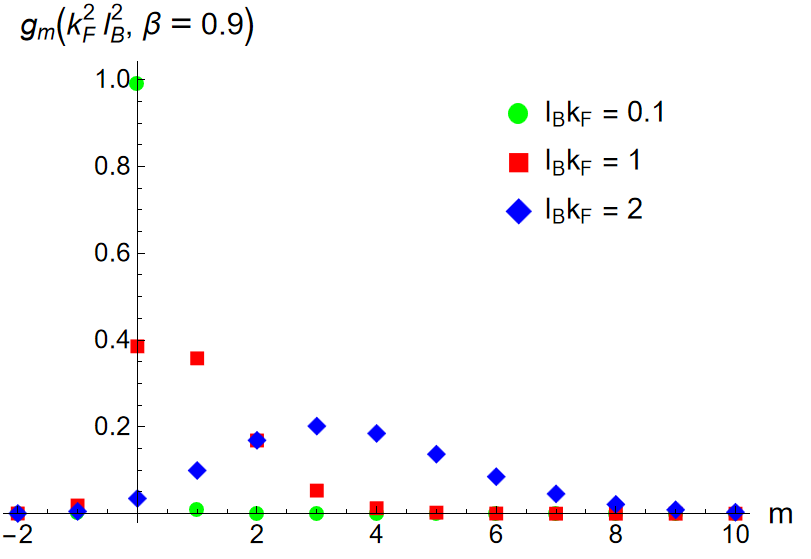}
    \end{subfigure}

    \vspace{0.5cm}

%================ Middle row (k_F l_B = 1) =================%

    \begin{subfigure}[b]{0.48\textwidth}
        \centering
        \includegraphics[width=\linewidth]{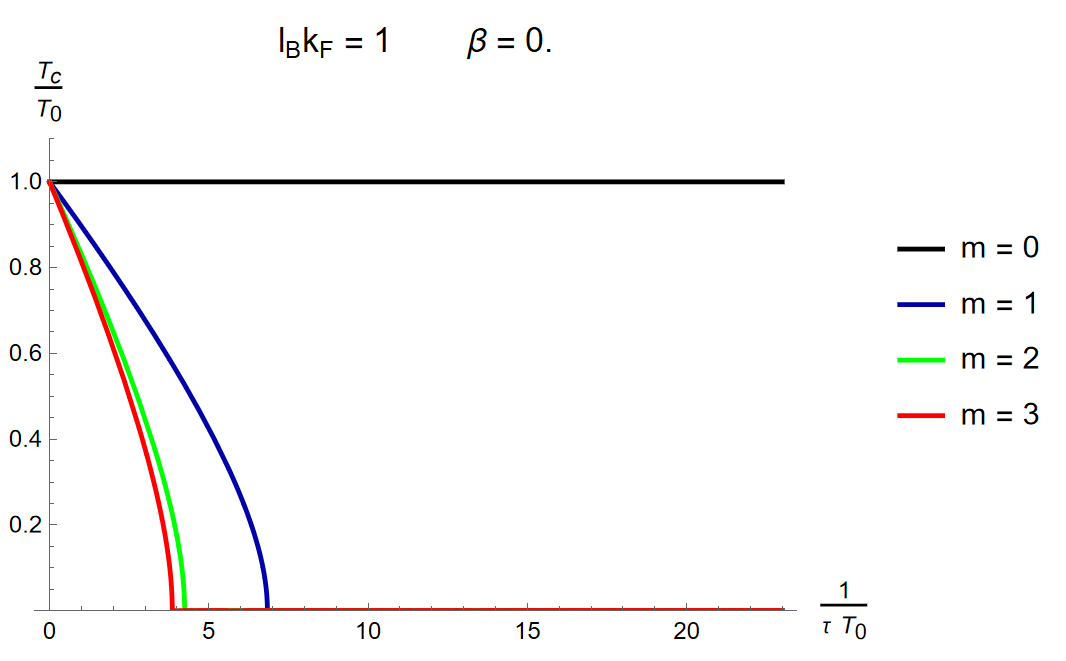}
    \end{subfigure}
    \hfill
    \begin{subfigure}[b]{0.48\textwidth}
        \centering
        \includegraphics[width=\linewidth]{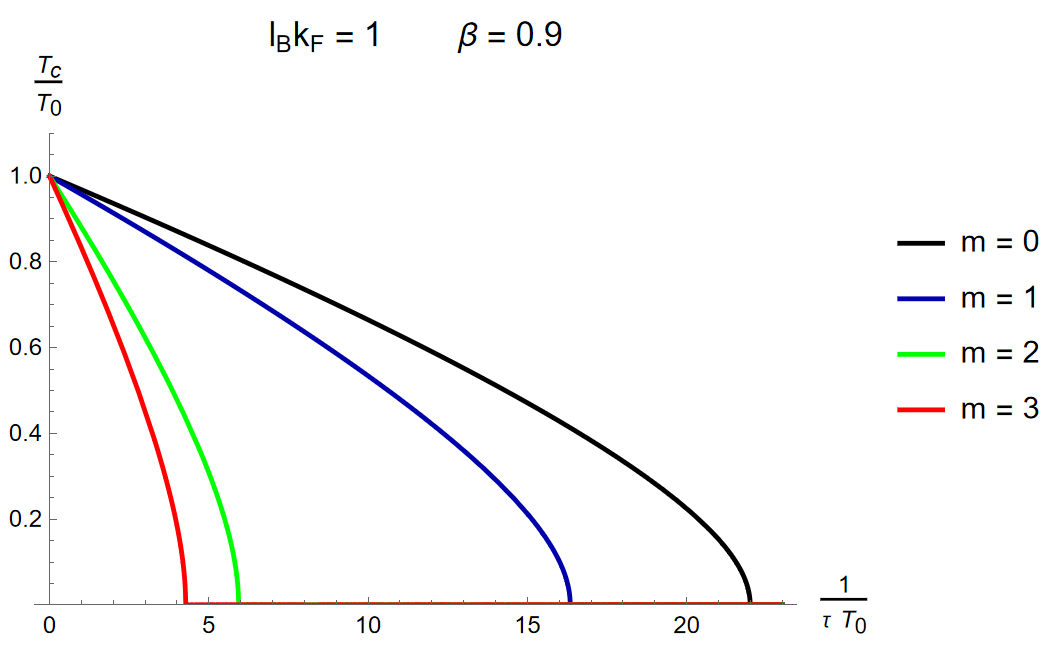}
    \end{subfigure}

    \vspace{0.5cm}

    %================ Bottom row =================%

    % Panel (c)
    \begin{subfigure}[b]{0.48\textwidth}
        \centering
        \includegraphics[width=\linewidth]{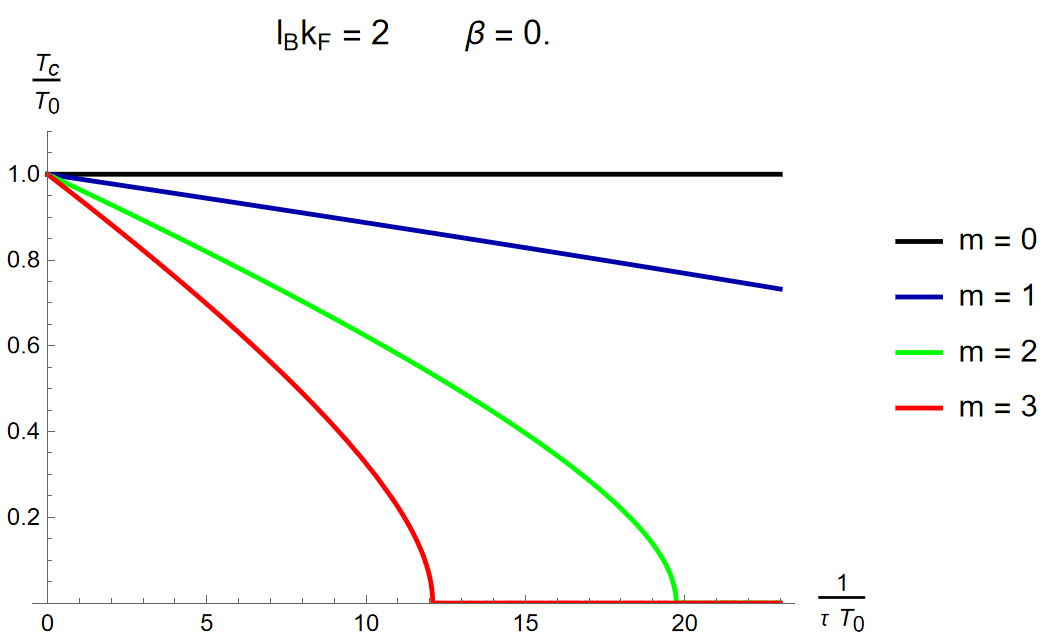}
    \end{subfigure}
    \hfill
    % Panel (d)
    \begin{subfigure}[b]{0.48\textwidth}
        \centering
        \includegraphics[width=\linewidth]{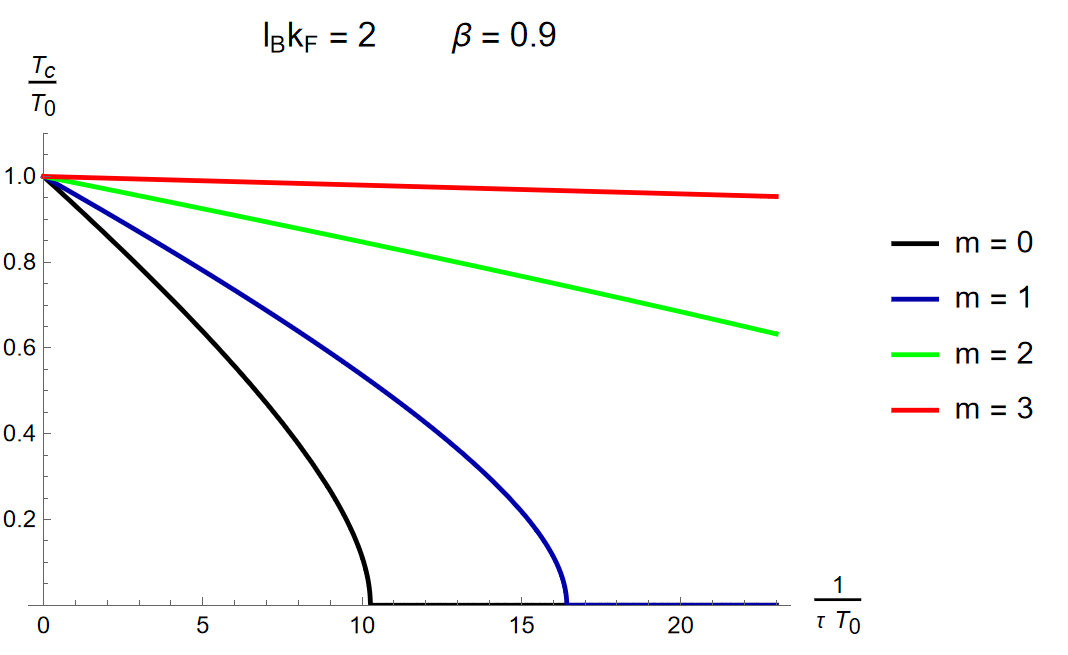}
    \end{subfigure}

    \caption{
    Top row: magnitude of $g_m((\ell_B k_F)^2,\beta)$ as a function of orbital angular momentum $m$ for different Berry curvature strength $\beta$. Middle  and bottom rows: critical temperature dependence on the scattering time $\tau$.
    }

    \label{fig:LLL data}
\end{figure*}

Here we consider a  form factor  
\begin{eqnarray}\label{LL form factor}
\Lambda_{\mathbf{k},\mathbf{p}} 
=
\exp\Bigg[
-\frac{\ell_B^{2}}{4} |\mathbf{k}-\mathbf{p}|^{2}
-i \beta \frac{ \ell_B^{2}}{2} (\mathbf{k} \times \mathbf{p})
\Bigg],
\end{eqnarray}
with $-1 \le \beta \le1$ parameterizing the strength of Berry curvature. For $\beta=\pm 1$ the above becomes exactly the Lowest Landau Level (LLL) form factor. The condition $|\beta| < 1$ ensures that the general inequality between the trace of the quantum metric and the Berry curvature is satisfied \cite{Roy_band_2014}.  From Eqns.~\eqref{fsym and fm definition} we have
\begin{eqnarray}
    f_{\text{sym}}=g_{0}(\ell^2_B k^2_F,0), \notag \\
    f_{m}=g_{m}(\ell^2_B k^2_F,\beta),
\end{eqnarray}
where function $g_{m}(a,\beta)$ is defined as
\begin{eqnarray}
    g_{m}(a,\beta)=e^{-a} \int e^{a(\cos\theta-i\beta \sin\theta)+im\theta}\frac{d\theta}{2\pi}= \notag \\
    =e^{-a} \left(\frac{1+\beta}{1-\beta}\right)^{\!m/2}  I_m(a\sqrt{1-\beta^2}),
\end{eqnarray}
 where $I_m$ is the modified Bessel function of the first kind. 
Numerical values of this function are presented in Fig.~\ref{fig:LLL data}. In the limit $k_F l_B \ll 1$ (when the band geometry is `weak') the usual results of Anderson (for $m = 0$) and Abrikosov-Gorkov (for $m \neq 0$) are recovered. For $\beta=0$ the channels that are most robust to disorder are typically those with small angular momentum $m$. However, as band geometry and Berry curvature effects become more pronounced with increasing $l_Bk_F$ and $\beta$, the most disorder-robust channel can shift to higher angular momenta. For example, in the bottom right panel of Fig.~\ref{fig:LLL data}, the most robust angular momentum becomes $m=3$ for $l_Bk_F=2$ and $\beta=0.9$. The Berry flux enclosed by the Fermi surface is $\Phi=\pi \beta (l_Bk_F)^2$, giving $\Phi=0.9\pi$ and $3.6\pi$ for $l_Bk_F=1$ and $2$ respectively, at $\beta=0.9$. 

In the opposite limit $k_F l_B \gg 1$ (`strong' band geometry), $g_m(\ell^2_B k^2_F,\beta)$ is a small quantity which varies slowly with m and in the presence of finite Berry curvature ($\beta \neq 0$) can have a maximum at nonzero angular momentum $m$, making this channel the most robust to disorder. This robustness can be understood as a strong suppression of large-momentum scattering by the band geometry. The effective pair breaking strength is controlled by  $f_{\text{sym}}-f_m$, and therefore decreases as $f_m$ approaches $f_{\text{sym}}$. In an extreme case when $f_{\text{sym}}=f_m$ that channel is completely unaffected by impurity scattering, which is an analog of Anderson's theorem.

\subsection{Sharply peaked Berry curvature: n-layer rhombohedral graphene}
Here we consider a toy two-band model for $n$-layer rhombohedral graphene in the presence of a weak displacement field. A convenient low-energy description is formulated in terms of the $A$-sites of the bottom graphene layer coupled to the $B$-sites of the top graphene layer. The displacement field adds a potential difference between these two sites. The single-particle Hamiltonian is 
\begin{eqnarray}\label{massive graphene}
    H(\boldsymbol{k}) =
    \begin{pmatrix}
    m_0 &  v (\,k_x + i k_y\,)^n \\[6pt]
     v (\,k_x - i k_y\,)^n & -m_0
    \end{pmatrix}\equiv \notag \\
    \begin{pmatrix}
    m_0 &  vk^n e^{in\theta_{\boldsymbol{k}}} \\[6pt]
     vk^n e^{-in\theta_{\boldsymbol{k}}} & -m_0
    \end{pmatrix}.
\end{eqnarray}
The corresponding eigenvalues are
\begin{eqnarray}
    E_\pm = \pm E_{\boldsymbol{k}}, \qquad E_{\boldsymbol{k}} = \sqrt{m_0^2 + (vk^n)^2}.
\end{eqnarray}
For $m_0>0$ a convenient choice of eigenvectors is:
\begin{align}
    &|\psi_+\rangle =
    \begin{pmatrix}
    \sqrt{\dfrac{E_{\boldsymbol{k}} + m_0}{2 E_{\boldsymbol{k}}}} \\[2pt]
    e^{-in\theta_{\boldsymbol{k}}}\sqrt{\dfrac{E_{\boldsymbol{k}} - m_0}{2 E_{\boldsymbol{k}}}}
    \end{pmatrix}, \notag \\
    &|\psi_-\rangle =
    \begin{pmatrix}
     - e^{in\theta_{\boldsymbol{k}}}\sqrt{\dfrac{E_{\boldsymbol{k}} - m_0}{2 E_{\boldsymbol{k}}}} \\[2pt]
    \sqrt{\dfrac{E_{\boldsymbol{k}} + m_0}{2 E_{\boldsymbol{k}}}}
    \end{pmatrix}.
\end{align}
If we now set the chemical potential to lie in the conduction band, forming a circular Fermi surface of momentum $k_F$, then the form factors near the Fermi surface can be written as 
\begin{align}\label{lambdas and s}   &\Lambda_{\boldsymbol{k},\boldsymbol{k}'}=\langle \psi_+(\boldsymbol{k}) |\psi_+(\boldsymbol{k}') \rangle= \notag \\
&=
    \frac{E_0+m_0}{2E_0}+\frac{E_0-m_0}{2E_0}e^{in(\theta_{\boldsymbol{k}}-\theta_{\boldsymbol{k}'})} ;\notag \\
    &E_0=\sqrt{m_0^2+(vk_F^n)^2}.
\end{align}
As we can see, this choice of eigenvectors produces rotationally invariant $\Lambda_{\boldsymbol{k},\boldsymbol{k}'}$, so assuming that the pairing interaction is also rotationally invariant, we can once again choose an order parameter of the form $\Delta_{\boldsymbol{k}}=\Delta e^{im\theta_{\boldsymbol{k}}}$. Also for this particular form factor we have the identity $\Lambda_{\boldsymbol{k},\boldsymbol{k}'}=\Lambda_{-\boldsymbol{k},-\boldsymbol{k}'}$. We then immediately find $f_{\text{sym}}$:
\begin{eqnarray}
    f_{\text{sym}}=\int |\Lambda_{\boldsymbol{k},\boldsymbol{p}}|^2 \frac{d\theta_{\boldsymbol{p}}}{2\pi}=\frac{E_0^2+m_0^2}{2E_0^2}.
\end{eqnarray}
and $f_m$ only has a finite number of nonzero angular harmonics:
\begin{eqnarray}
    f_m
    &=&
    \int \Lambda_{\boldsymbol{k},\boldsymbol{p}}^2 e^{im(\theta_{\boldsymbol{p}}-\theta_{\boldsymbol{k}})}
    \frac{d\theta_p}{2\pi}=
    \nonumber\\
    &=&
    \begin{cases}
    (\frac{E_0+m_0}{2E_0})^2, & m=0,\\
    \frac{E_0^2-m_0^2}{2E_0^2}, & m=n,\\
    (\frac{E_0-m_0}{2E_0})^2, & m=2n,\\
    0, & \text{otherwise}.
    \end{cases}
\end{eqnarray}
In the limit $m_0 \gg v k_F^n$ we effectively have a band with $\Lambda_{\boldsymbol{k},\boldsymbol{k}'}=1$ and thus  $f_{\text{sym}}=1$, $f_m=\delta_{m,0}$, which means no suppression of $T_c$ for an s-wave superconductor (Anderson's theorem) and a regular Abrikosov-Gorkov formula for every other pairing channel.\par 
In the opposite limit of $m_0 \ll v k_F^n$ the Berry curvature is concentrated near $k=0$ and the parameters become
\begin{align}
    &f_{\text{sym}}=\frac{1}{2}, \notag \\
    &f_m
    =
    \begin{cases}
    \frac{1}{4}, & m=0,\\
    \frac{1}{2}, & m=n,\\
    \frac{1}{4}, & m=2n,\\
    0, & \text{otherwise}.
    \end{cases}
\end{align}

\begin{figure}[t]
    \centering
    \includegraphics[width=0.48\textwidth]{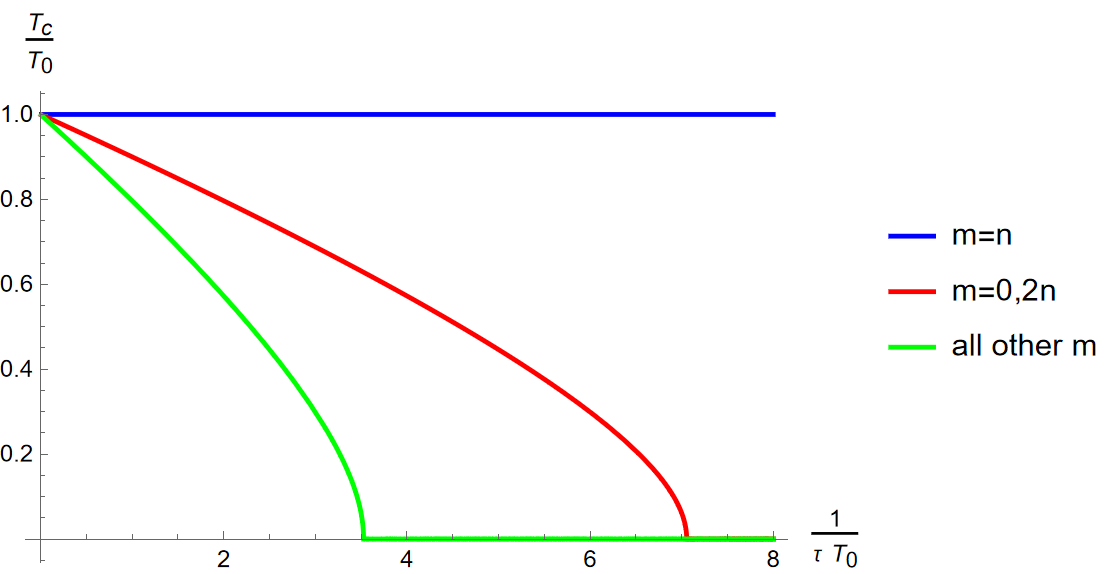}
    \caption{Dependence of superconducting critical temperature $T_c$ on scattering time $\tau$ for electron-doped n-layer rhombohedral graphene.}
    \label{fig:massive graphene}
\end{figure}

The suppression of superconductivity with disorder for small $m_0$ is demonstrated in Fig.~\ref{fig:massive graphene}. As we can see, $m=n$ is not suppressed at all, $m=0$ and $m=2n$ are weakly suppressed and all other order parameters follow a universal strongly suppressed curve.

\section{$T_c$ suppression in terms of residual resistivity}
We now express the mean field $T_c$ suppression due to disorder in terms of the residual resistivity rather than the scattering time $\tau$ as the former is an easily measured quantity.

\par 
The residual resistance can be extracted from the Drude formula, which for arbitrary rotationally symmetric dispersion of spinless fermions takes the form:
\begin{eqnarray}
    \rho=\frac{4\pi}{e^2v_Fk_F\tau_{\text{tr}}},
\end{eqnarray}
where $v_F$ and $k_F$ are Fermi velocity and Fermi momentum respectively.
Since we take the effects of band geometry into account, the transport time is also affected by it. Assuming that the scattering from impurity is isotropic, Fermi's golden rule implies that transport scattering time takes the form 
\begin{align}\label{ftr}
    &\frac{1}{\tau_{\text{tr}}}=\frac{f_{\text{tr}}}{\tau} \notag \\
    &f_{\text{tr}}=\int |\Lambda_{\boldsymbol{k},\boldsymbol{p}}|^2(1-\cos(\theta_{\boldsymbol{k}}-\theta_{\boldsymbol{p}})) \frac{d\theta_{\boldsymbol{p}}}{2\pi}\frac{d\theta_{\boldsymbol{k}}}{2\pi}.
\end{align}
Just as in the case of $f_{\text{sym}}$, the quantity $f_{\text{tr}}$ may depend on $\boldsymbol{k}$, meaning that the scattering time is different for different momenta, but we neglect this dependence. \par
We can now express $\frac{1}{\tau}$ through $\rho$ and substitute it into Eq. \eqref{crit temp eqn with disorder}, which then takes the form:
\begin{align}\label{Tc over rho}
    &\ln \left(\frac{T_0}{T_c}\right)= \notag \\
    & \psi\left(\frac{1}{2}+\frac{f_{\text{sym}}-f_m}{8\pi f_{\text{tr}}}\frac{v_F k_F}{E_F}\frac{E_F \rho}{T_0 R_0} \frac{T_0}{T_c}\right)-\psi\left(\frac{1}{2}\right).
\end{align}
Here $E_F$ is Fermi energy and $R_0=\frac{h}{e^2}\approx 25.8k\Omega$ is the resistance quantum. The fraction $\frac{v_F k_F}{E_F}$ for dispersion $\varepsilon_k \propto k^\alpha$ is equal to $\alpha$, so this factor is expected to be of order 1 for most bands.
\par

% Here we assumed that $m>0$ so we put the $e^{-i\theta_k}$ factor in front of $\sqrt{\dfrac{E_0 - m}{2 E_0}}$, so the components of wavevectors behave analytically near $\boldsymbol{k}=\boldsymbol{0}$. 

\section{A realistic calculation: an application to twisted MoTe$_2$}
In previous sections we have seen that the effects of disorder  on the mean field critical temperature  $T_c$ depend on the band geometry. In this section we apply this understanding to the observation of chiral superconductivity next to the $2/3$ FQAH state in t$\text{MoTe}_2$ \cite{Xu2025_SCdopeTMD}. As explained in the introduction, an interesting explanation of the superconductivity is that it originates from a mobile fluid of anyons created by doping the FQAH state \cite{Shi2024_doping, Shi2025_anyon_delocalization, Nosov2025_plateau}. Here, however, we consider the alternate possibility, suggested by some authors, that the superconductivity is a BCS (Kohn-Luttinger-type) instability of a `large' Fermi surface obtained by doping holes into the filled $\nu = 1$ Chern insulator \cite{xu2025chiral,Guerci_from_2025,guerci2026topological}. To that end we study a somewhat realistic model for the band structure of t$\text{MoTe}_2$ at the pertinent twist angles.  We will see that the large measured resistivities put strong constraints on the interpretation in terms of a BCS instability of the large hole Fermi surface. 

%the suppression of $T_c$ to the measured residual resistivity In order to experimentally check that dependence one needs to measure $T_c$ as a function of scattering time. Here we propose the way to measure scattering time through residual resistance (a resistance obtained by extrapolating the normal state $\rho(T)$ to zero temperature).

Eqn.~\eqref{Tc over rho} expresses the mean field critical temperature in terms of the residual resistivity $\rho$ through only three real parameters $f_m$, $f_{\text{sym}}$ and $f_{\text{tr}}$ which are extracted from the band form factors $\Lambda_{\boldsymbol{k},\boldsymbol{p}}$. In order to demonstrate this dependence, we now do a calculation for twisted bilayer MoTe$_2$ using the reasonably realistic  skyrmion crystal model for the band structure introduced in \cite{Wu2019}, which has a Hamiltonian for electrons in the K valley
\begin{eqnarray}
H =
\begin{pmatrix}
-\frac{\hbar^2(\boldsymbol{k}-\boldsymbol{\kappa}_+)^2}{2m^*} + V_1(\boldsymbol{r})
&
T(\boldsymbol{r})
\\[6pt]
T^\dagger(\boldsymbol{r})
&
-\frac{\hbar^2(\boldsymbol{k}-\boldsymbol{\kappa}_-)^2}{2m^*} + V_{-1}(\boldsymbol{r})
\end{pmatrix}.
\end{eqnarray}
Here $\boldsymbol{\kappa}_{\pm}$ are basis vectors located at the edges of moir\'e Brillouin zone, $m^*$ is the effective mass and layer potentials $V_{l}(\boldsymbol{r})$ and tunneling elements $T(\boldsymbol{r})$ are defined as follows:
\begin{align}
&V_l(\boldsymbol{r})
=
2V \sum_{j=1,3,5}
\cos(\mathbf{G}_j\cdot \boldsymbol{r} + l\psi),
\quad l=\pm 1
\notag \\
&T(\boldsymbol{r})
=
w\left(1 + e^{-i\mathbf{G}_2\cdot \boldsymbol{r}}
+ e^{-i\mathbf{G}_3\cdot \boldsymbol{r}}\right).
\end{align}

In the equations above $\mathbf{G}_i =
4\pi/(\sqrt{3}\,a_M)
\Big(
\cos\!\left[(i-1)\pi/3\right],
\;
\sin\!\left[(i-1)\pi/3\right]
\Big)$ are moir\'e reciprocal lattice vectors, $a_M\approx a_0/\theta$ is the moiré lattice constant. For this particular calculation we take the following parameters: $(V,\psi,w,\theta) \approx (8~\mathrm{meV}, -89.6^\circ, -8.5~\mathrm{meV},3.83^\circ)$. \par 

\begin{figure}[t]
    \centering
    \includegraphics[width=0.48\textwidth]{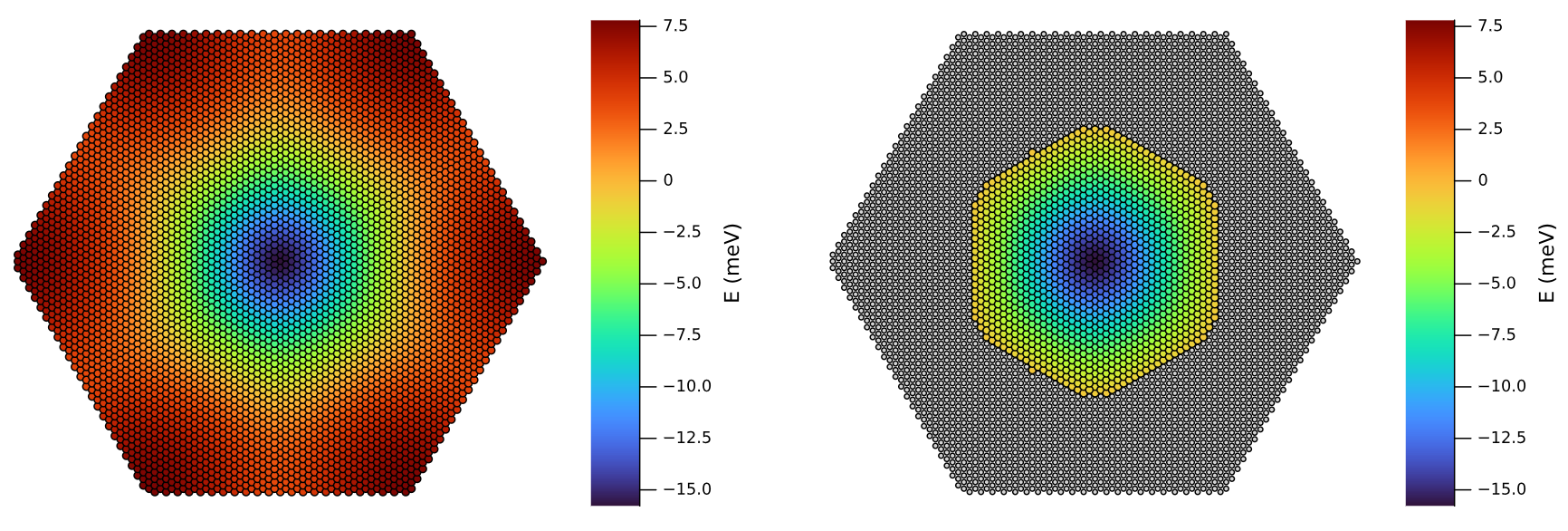}
    \caption{Left panel: band dispersion of the first valence band of 3.83$^\circ$ twisted MoTe$_2$. Right panel:  filled states of the first valence band at $\nu=0.72$ hole doping. Empty states are shown with grey and electrons fill the region near Gamma point.}
    \label{fig:dispersion}
\end{figure}

Motivated by experiment \cite{Xu2025_SCdopeTMD}, we focus on the hole doping of approximately $\nu_h \approx 0.72$ for which the superconductivity is observed. The dispersion of the first valence band and the filled electron states at that filling are shown in Fig.~\ref{fig:dispersion}. As we can see,  despite the absence of exact $C_6$ symmetry the Fermi surface is almost circular. Since the anisotropy is rather small, we neglect it and in all subsequent calculations we use quantities averaged over the Fermi surface. For a general discussion of the effects of anisotropy, see Ref.~\cite{Sedov_quantum_2025}. 

\par
\begin{figure}[t]
    \centering

    \begin{minipage}{0.5\textwidth}
        \centering
        \includegraphics[width=\textwidth]{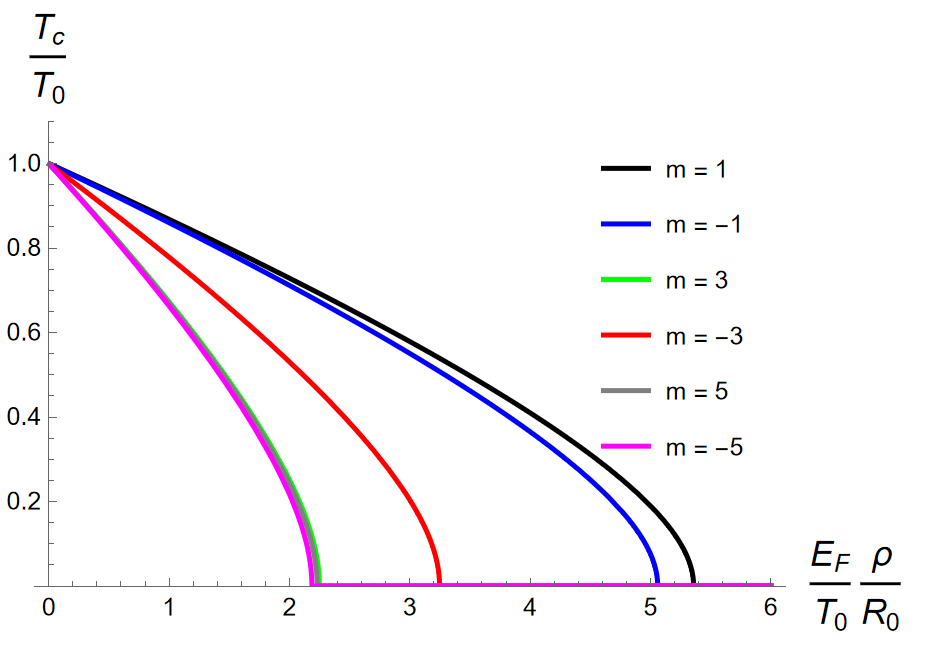}
        \vspace{2mm}
        
    \end{minipage}
    \hfill
    
\begin{minipage}{0.36\textwidth}
\raggedright
\setlength{\tabcolsep}{3pt}
\renewcommand{\arraystretch}{1.2}

\begin{tabular}{ccccccccc}
\toprule
$f_{\mathrm{sym}}$ & $f_{\mathrm{tr}}$ & $f_{1}$ & $f_{-1}$ & $f_{3}$ & $f_{-3}$ & $f_{5}$ & $f_{-5}$ \\
\midrule
0.305 & 0.163 & 0.183 & 0.176 & 0.014 & 0.104 & 0.012 & 0.006 \\
\bottomrule
\end{tabular}

\end{minipage}

    \caption{
    Top panel: superconducting critical temperature dependence on residual resistivity $\rho$ for $3.83^\circ$ twisted MoTe$_2$.  Bottom panel: values of function $f_m$ for hole doping $\nu_h=0.72$ and different angular momenta $m$.}

    \label{fig:tMoTe2_Tc_over_rho}
\end{figure}

Now we numerically evaluate $\Lambda_{\boldsymbol{k},\boldsymbol{p}}$. We choose a smooth periodic gauge by parallel transporting the Bloch vector along the Fermi surface and uniformly distributing the accumulated Berry phase. Then using equations \eqref{fsym and fm definition}, \eqref{ftr} and \eqref{Tc over rho}, we can find the dependence of critical temperature  on resistance. The numerical results are presented in Fig.~\ref{fig:tMoTe2_Tc_over_rho}. The small values of $f_{\text{sym}}$ and $f_{\text{tr}}$ reflect the strong suppression of impurity scattering by the Bloch-state overlaps. Since we consider a spin- and valley-polarized system, we will only look at odd angular momentum channels. The $m=\pm1$ channels are rather close to each other, having $\frac{f_m}{f_{\text{sym}}} \approx 0.6$, meaning that both of these channels are very robust to disorder. A substantial difference between two opposite order parameter phase windings can be seen for $m=\pm 3$. While $f_{-3}$ is rather large, $f_{3}$ is tiny and is of the same order as $f_5$. We suggest that this is a feature of a particular Berry phase winding over the Fermi surface. Lastly, $f_{\pm5}$ are so small that the curve is mostly defined by $f_{\text{sym}}$ and $f_{\text{tr}}$.

As we can see, according to Eq.~\eqref{Tc over rho} there is a critical residual resistance for which the superconductivity vanishes. In order for superconductivity to survive the residual resistivity should satisfy the inequality:
\begin{eqnarray}
    \rho \leq (2\pi e^{-\gamma}) \frac{E_F}{v_F k_F}\frac{f_{\text{tr}}}{f_{\text{sym}}-f_{\text{m}}} R_0 \frac{T_0}{E_F}.
\end{eqnarray}
For the hole doping $\nu\approx 0.72$ of $3.83^\circ$ tMoTe$_2$ the fraction $\frac{E_F}{v_F k_F}$ using the averaged Fermi velocity and momentum is approximately equal to $1.137$. Then, based on the table in Fig.~\ref{fig:tMoTe2_Tc_over_rho} and assuming that a weakly coupled BCS superconductor has $\frac{T_0}{E_F}<0.1$ we can write an estimate for the residual resistivity:
\begin{eqnarray}
    \rho \leq\begin{cases}
    13.8k\Omega, & m=1,\\
    13.1k\Omega, & m=-1,\\
    5.81k\Omega, & m=3,\\
    8.39k\Omega, & m=-3,\\
    5.76k\Omega, & m=5,\\
    5.65k\Omega, & m=-5.\\
    \end{cases}
\end{eqnarray}
In the absence of band geometry effects where all form factors are equal to one, the critical residual resistivity would be the same for all nonzero angular momentum channels and equal to $\rho_c\approx 10.35k\Omega$. Although band geometry always decreases the suppression of the critical temperature with disorder (since $f_{\text{sym}}-f_m<1$), the critical residual resistivity can be either increased or decreased, as band geometry also changes the transport scattering rate by a factor $f_{\text{tr}}$.
\par
As we can see, the residual resistivities obtained above are roughly in the same range as that inferred from the measurements in Ref. ~\cite{Xu2025_SCdopeTMD}. The studies that suggest a BCS pairing mechanism for this material \cite{xu2025chiral,Guerci_from_2025,guerci2026topological} argue that the leading superconducting instability has f-wave symmetry. Since the upper bounds for $m=\pm3$ are rather close, our estimate is relatively insensitive to the particular chirality of the f-wave state, providing a direct criterion that can be compared to an experimentally measured residual resistivity. We thus find that the large residual resistivity in Ref. ~\cite{Xu2025_SCdopeTMD} is in tension with a BCS interpretation of the observed superconductivity in tMoTe$_2$.

\section{Spontaneous disorder-induced vortices}
In previous sections we considered a mean-field treatment of disordered superconductivity, where the order parameter is assumed to be homogeneous and the gap equation is averaged over disorder. In this section we consider other mechanisms of superconductivity suppression which go beyond mean-field. For a chiral time-reversal-breaking superconductor stabilized by a BCS mechanism, we pose the question of whether a random potential induces spontaneous nucleation of vortices, resulting in an `Anomalous Vortex Glass' (AVG) phase \cite{Shi2025_anyon_delocalization}. Such an AVG has been argued to be induced within anyonic mechanisms for superconductivity near the doped Fractional Quantum Anomalous Hall state \cite{Shi2025_anyon_delocalization,senthil2026fractionalized}. Thus it is of interest to ask whether an AVG is also expected within the BCS mechanism of interest in the present paper. 

% It is well-known that in two spatial dimensions superconductivity may be unstable to vortex-antivortex pair formation, which eventually leads to a Berezinskii Kosterlitz Thouless (BKT) transition \cite{Kosterlitz_ordering_1973}. Usually, when we consider vortex formation the details of band structure are encoded in Ginzburg-Landau (GL) parameters and they do not affect the overall physical picture. However, when the underlying band has an orbital magnetization, the disorder may affect the electronic density and thus the spatial variation of magnetization, which will couple to magnetic field and vorticity.
 This question is particularly relevant in the setting of BCS superconductivity in moir\'e  MoTe$_2$ bilayers where the clean BCS superconducting states have been argued to be analogous to 2D Abrikosov vortex lattices in the effective periodic magnetic field created by the moir\'e Skyrmion texture \cite{guerci2026topological}. The Abrikosov vortex lattice of ordinary 2D superconductors in a magnetic field is expected to become unstable towards a vortex glass under weak disorder \cite{Fisher_vortex-glass_1989, Fisher_thermal_1991}. The fact that this instability appears for weak disorder is a consequence of viewing the Abrikosov lattice as breaking spontaneously the translational symmetry, which is a reasonable approximation in the BCS setting when the coherence lengths and thus vortex sizes are much larger than inter-atomic distances, allowing for an accurate  description in terms of a continuum model. But the periodic effective magnetic field that creates the Abrikosov-like lattice in the BCS superconductors proposed for twisted MoTe$_2$, would have, by construction, one flux per moir\'e unit cell (i.e. two effective superconducting flux quanta), and therefore, there is generically a simple commensuration of vortices with moir\'e unit cells, that generically is expected to lead to vortex pinning within moir\'e unit cell. This pinning of the vortices prevents weak disorder from leading into a vortex glass, since the vortex positions remain adiabatically pinned to their moir\'e unit cells for weak disorder potentials. In this section, we will address whether weak disorder is able to nucleate additional vortex defects on top of such underlying ordered vortex-like lattice of the MoTe$_2$ within an effective long-wavelength description that assumes a coarse graining of the superconductivity over length scales larger than the moir\'e unit cell. Since the underlying intra-unit cell vortices can be viewed as pinned Coulomb charges which are neutralized by the periodic effective skyrmion magnetic field background, they are an effective vortex neutral background which disappears upon coarse graining. Within this point of view, the key question we are faced with is if disorder which is consistent with time-reversal symmetry breaking of the moir\'e Chern band would be able to nucleate additional vortex defects on this ordered background.

To address this question, we consider a long-wavelength  Ginzburg-Landau (GL)  theory of a disordered chiral superconductor. A new feature generically present in the chiral topological bands of the normal state is an orbital magnetization which can couple both to the magnetization of the SC and, due to the broken time-reversal symmetry, to a random potential.  The  orbital magnetization is described by a spatially dependent vector $\mathbf{M}$ pointing perpendicular to the plane. The resulting functional is
\begin{eqnarray}
    f=\alpha|\psi^2|+\frac{\beta}{2}|\psi|^4+K|(\mathbf{\nabla}-\frac{2e}{c}\mathbf{A})\psi|^2 +\notag \\
     +\gamma_M V(r)M(r)+\alpha_M (M(r)-M_0)^2 \notag \\
   \chi_0 M(r) \nabla \times j(r)+\gamma V(r)|\psi^2|.
\end{eqnarray}
Here $\psi$ is the usual superconducting order parameter, $j$ is the supercurrent associated with $\psi$, and $V(r)$ is a random potential. The terms in the second line can be simplified by taking an approximation that the spatially varying part of the  orbital magnetization is always proportional to the impurity potential:
\begin{eqnarray}
    (M(r)-M_0)\propto V(r).
\end{eqnarray}
Substituting this into the GL energy functional and omitting the total derivative term $\chi_0 M_0 \nabla \times j(r)$, we obtain: 
\begin{eqnarray}\label{GL free eng}
    f=\alpha|\psi^2|+\frac{\beta}{2}|\psi|^4+K|(\mathbf{\nabla}-\frac{2e}{c}\mathbf{A})\psi|^2 +\notag \\
     +\gamma V(r)|\psi^2|+ \chi V(r) \nabla \times j(r) .
\end{eqnarray}
The $\gamma$ term is a local modulation of $\alpha$ in the disordered system, and plays no significant role in questions of vortex nucleation. The $\chi$ term, however, is important: in the presence of a randomly varying orbital magnetization, it describes a tendency of disorder to induce local vorticity. Without orbital magnetization, the disorder would simply locally enhance or suppress superconductivity depending on the sign of the impurity potential, but the phase of the order parameter would be homogeneous throughout the whole sample. With orbital magnetization there are two competing effects: first, creating spatial phase variations costs energy, resulting in the interaction term for vortices. Second, a vortex may gain energy after being pinned to the impurity potential due to orbital magnetization.

The vorticity is localized near the vortex cores. For vortices centered at positions $\boldsymbol{r}_i$, $\boldsymbol{\nabla} \times \boldsymbol{j}(\boldsymbol{r})\propto\sum_i q_i\delta(\boldsymbol{r}-\boldsymbol{r}_i)$, so that the last term in Eq.~\eqref{GL free eng} generates a random potential $v(\boldsymbol{r}_i)$ coupled directly to vortex charge. This physics is captured by 
considering the energy of a finite concentration of $N$ vortices and $N$ antivortices. The equality of their numbers ensures zero net vorticity and avoids energy cost that grows logarithmically with system size. The resulting effective Coulomb gas Hamiltonian is:
\begin{equation}\label{Coulomb gas}
    H=-\sum_{i>j} q_i q_j U \ln \frac{|r_i-r_j|}{\xi} +\sum_i v(r_i) q_i +2N E_{\text{core}}.
\end{equation}
Here $\xi$ is a coherence length, $U=4\pi K\frac{|\alpha|}{\beta}=2\pi \rho_s$ is the strength of interaction between vortices, expressed in terms of superconducting phase stiffness $\rho_s$, $v(r_i)$ is a renormalized impurity potential term coupling to the vorticity,  and $q_i=\pm 1$ for a vortex and an antivortex respectively. 

{\it Naively}, the random vortex potential does not lead to nucleation of vortex-antivortex pairs. This is because the typical energy gained by nucleating a vortex-antivortex pair separated by a distance $r$ is $O(\sigma)$ where $\sigma$ is the width of the probability distribution of $v$ which competes unfavorably with the stiffness $\ln(r)$ energy cost. 
  However, this does not take into account rare-region effects coming from tails of the probability distribution of $v$. 

Let us take $v(r)$ to be drawn from a Gaussian random potential with zero mean and standard deviation $\sigma$:
\begin{eqnarray}
    v(r)\in \mathcal{N}(0,\sigma).
\end{eqnarray}
Here $x \in \mathcal{N}(x_0,\sigma)$ denotes a random variable with distribution 
\begin{eqnarray}
    p(x)=\frac{1}{\sqrt{2\pi \sigma^2}} e^{-\frac{(x-x_0)^2}{2\sigma^2}}.
\end{eqnarray}
We also assume that the correlation function of this potential decays quickly when the distance exceeds the coherence length $\xi$, so that the potentials at two points separated by a distance larger than $\xi$ can be treated as independent.  

Let us first focus on a single vortex-antivortex pair. We can parametrize its position by a center-of-mass coordinate $\boldsymbol{R}$ and a relative separation $\boldsymbol{r}$
\begin{align}\label{total vortex energy}
    E=&U\ln \left(\frac{r}{\xi}\right)+2E_{\rm core}+\notag \\
    & v\left(\boldsymbol{R}+\frac{\boldsymbol{r}}{2}\right)-v\left(\boldsymbol{R}-\frac{\boldsymbol{r}}{2}\right).
\end{align}
Suppose the magnitude of vector $\boldsymbol{r}$ lies in the range from $r$ to $r+\Delta r$ and we are trying to optimize the energy over the center-of-mass coordinate and the angular variable. Since the potentials at two vectors $\boldsymbol{r}$ become uncorrelated once their separation is larger than $\xi$, there are roughly $\frac{2\pi r \Delta r}{\xi^2}$ ways to choose $\boldsymbol{r}$ and after taking into account the center-of-mass coordinate, we see that there are approximately $\frac{2\pi r \Delta r S}{\xi^4}$ independent configurations, where $S$ is the area of the system. \par
The difference of two potentials for each such configuration is an independent random variable:
\begin{eqnarray}\label{def VR}
v\left(\boldsymbol{R}+\frac{\boldsymbol{r}}{2}\right)-v\left(\boldsymbol{R}-\frac{\boldsymbol{r}}{2}\right)\in\mathcal{N}(0,\sqrt{2}\sigma).
\end{eqnarray}
If we want the vortex-antivortex pair to be spontaneously induced even at zero temperature, the total energy from equation \eqref{total vortex energy} should be negative, or equivalently, the above random variable should be smaller than the following quantity:
\begin{eqnarray}
    E_{p}(r)=-U\ln (\frac{r}{\xi})-2E_{\rm core}.
\end{eqnarray}
Since our random variable is Gaussian, this probability   is given by an error integral:
\begin{eqnarray}
    p(r)=\int_{-\infty}^{E_{p}(r)} \frac{1}{\sqrt{4\pi \sigma^2}}e^{-\frac{E^2}{4\sigma^2}} dE.
\end{eqnarray}
If $|E_p(r)|$ is much larger than $\sigma$, which is the case for weak disorder, then we can approximate this by 
\begin{eqnarray}
    p(r)\approx \frac{\sigma}{\sqrt{\pi}|E_p(r)|} e^{-\frac{E_p^2(r)}{4\sigma^2}}.
\end{eqnarray}
Now if we are interested in the number of vortex-antivortex pair configurations with arbitrary r, we get
\begin{eqnarray}
    N_{\rm va}=S \int_{\xi}^{\infty}p(r) \frac{2\pi r dr}{\xi^4}.
\end{eqnarray}
Here we used $\xi$ as a natural short distance cutoff. If we now assume that we are in a dilute regime where the interaction between different vortex-antivortex pairs can be neglected, then we get a result for the vortex-antivortex pair concentration at zero temperature:
\begin{align}
    &\frac{N_{va}}{S}=\frac{2\sqrt{\pi}}{\xi^2}\frac{\sigma}{U}\notag \\ &\int_{1}^{\infty} \exp\Big[{-\frac{U^2}{4\sigma^2}\left(\ln x +\frac{2E_{\rm core}}{U}\right)^2} \Big] \frac{x dx}{\ln x+\frac{2E_{core}}{U}}.
\end{align}
As $\sigma \ll U,E_{\rm core}$, this integral can be approximated to be 
\begin{eqnarray}
    \frac{N_{va}}{S}\propto\frac{1}{\xi^2}\frac{\sigma}{U}\frac{\sigma^2}{E_{\rm core}^2}e^{-\frac{E_{\rm core}^2}{\sigma^2}}\propto\frac{1}{\xi^2}\frac{\sigma^3}{\rho_{s}^3}e^{-C^2\frac{\rho_{s}^2}{\sigma^2}},
\end{eqnarray}
where $C=\frac{E_{\rm core}}{\rho_s}$ is a numerical coefficient measuring the value of the core energy in terms of superfluid stiffness. So, in principle, there is a disorder-induced finite (though extremely small) concentration of vortex-antivortex pairs in the ground state. 

What role do these vortex-antivortex pairs have on the fate of long-range order at $T = 0$? Each vortex-antivortex pair acts as a random dipole in the Coulomb gas model. Thus we need to supplement our original random Coulomb gas by adding an additional random dipole field $\boldsymbol{P}(\boldsymbol{r})$ which we can take to be a Gaussian random field with short-range correlations: 
\begin{equation} 
\overline{P_\alpha(\boldsymbol{r}) P_\beta(\boldsymbol{r}')} = w^2 \delta_{\alpha\beta}\delta^{2}(\boldsymbol{r} -\boldsymbol{r'}) ,
\end{equation} 
 where $\alpha, \beta$ are the spatial components of the vector $\boldsymbol{P}$, and $w$ is a parameter denoting the strength of the induced dipolar disorder. Since the vortex-antivortex pair concentration is exponentially small, the dipole strength is much smaller than all other energy scales: $w\ll \sigma \ll \rho_s$.
We also ignore the correlations between the distributions of $\boldsymbol{P}(\boldsymbol r)$ and $v(\boldsymbol{r})$. 
 
Through the Poisson equation, the random dipoles then modify the effective vortex potential to 
\begin{eqnarray} 
v_{\rm eff} & = & v + \chi, \\
-\nabla^2\chi & = & 2\pi \boldsymbol{\nabla} \cdot \boldsymbol{P} ,
\end{eqnarray} 
$\chi$ is a Gaussian random field with zero mean. Its variance is readily stated in Fourier space: 
\begin{equation} 
\overline{|\chi(k)|^2}  = \frac{4\pi^2 w^2}{k^2} .
\end{equation} 
Thus we generate (at very long length scales) a vortex potential with long-range correlations in real space. This is identical to the vortex potential in the celebrated random phase shift 2d XY model with 
\begin{equation} 
\label{randomxy} 
H = -J \sum_{ij} \cos(\theta_i - \theta_j - a_{ij}) ,
\end{equation} 
with $a_{ij}$ taken to be independent random variables on each bond \cite{Korshunov_2006}. We can now use known results on this model. Specifically, it has been shown that at $T = 0$, there is a critical disorder strength $w \sim \rho_s$ needed to destroy the XY long-range order. As we are working in the limit $w \ll \rho_s$, we conclude that long-range order survives in our system at $T = 0$. Further, it is known that the quasi-long-range order at non-zero $T$ survives with some weak glassy effects, and that eventually there is a sharp finite-temperature phase transition.

Thus we conclude for the chiral superconductor, a weak random potential does not destroy the $T = 0$ long-range order or the associated finite-$T$ phase transition. This conclusion is expected  whenever the superconductor is obtained via the usual BCS pairing mechanism of a parent electronic Fermi-surface, since this BCS point of view does not provide a natural mechanism for softening of the superconducting phase stiffness. In contrast, if the mechanism for obtaining the superconducting state is through the creation of a mobile anyon fluid, the phase stiffness will be small at small doping away from the FQAH state\cite{Shi2024_doping,Shi2025_anyon_delocalization}, and this leads to more dramatic effects of a random potential of fixed strength, as we review next.

\section{Contrast with anyon-induced superconductivity} 
An alternate interpretation of the superconductivity observed in tMoTe$_2$ near the $2/3$ FQAH is that it is obtained as the ground state of a mobile fluid of anyons rather than through the BCS pairing instability of an electronic Fermi surface. Here we briefly discuss the answers to the two questions of interest in this paper within the anyon fluid framework. At the outset we note that there are many different routes through which an anyon fluid can yield a superconductor upon doping the $2/3$ FQAH state \cite{Shi2024_doping,Zhang2025_SU(3)1dope, Shi2025_anyon_delocalization,  Nosov2025_plateau,Pichler2025_anyonSC,Shi2025_nonAbelian_TSC,lotrivc2026phases,Fan2026_weakpairing_SC,Wang2026_U3anyonSC,shi2026superconductivity,senthil2026fractionalized,han2026single,zhang2026color,mehta2026coloring}. For a recent discussion of some of the possibilities, see Ref. ~\cite{shi2026superconductivity}. 

The first question is the relationship between the highly resistive metallic normal state and the superconductivity. Very recent theoretical work \cite{senthil2026fractionalized,han2026single,shi2026superconductivity} has proposed that this metallic state, obtained in a fluid of charge $1/3$ anyons, is fractionalized: it has a Fermi surface of sharply defined charge-1/3 {\it fermionic} quasiparticles that are coupled to a (twisted) $Z_3$ gauge field. This state - denoted a $Z_3$ Orthogonal Metal \cite{nandkishore2012orthogonal} - is a non-Fermi liquid in the sense of not having sharp electronic quasiparticles but has a simple low energy description in terms of the Fermi surface excitations of the fractionalized fermions. Pairing these fractionalized fermions leads to a charge-$2e$ superconductor with no coexisting topological order. In this route the high resistance of the normal state is attributed \cite{senthil2026fractionalized} to the small charge of the fermionic quasiparticles even when their mean free path is large compared to their Fermi momentum. The limitations coming from the (modified) Anderson/Abrikosov-Gorkov theory of dirty superconductors are evaded in this mechanism as the impurity scattering rate is small despite the large resistance. 

The second question is about the stability of the  chiral superconducting state to vortex nucleation in a random potential. In many anyon-induced mechanisms (including the one involving the $Z_3$ Orthogonal Metal), it has been proposed \cite{Shi2025_anyon_delocalization,senthil2026fractionalized} that the first superconductor that forms, in the presence of disorder, is an Anomalous Vortex Glass with no ordinary long-range XY order at $T = 0$ (though an Edwards-Anderson type order parameter describing the frozen phase is present)  \cite{Shi2025_anyon_delocalization}. Correspondingly, there is no BKT transition at non-zero $T$. Rather, through thermal creep of vortices, the state at any non-zero $T$ becomes resistive. 

From the point of view of random $XY$ models, what is the difference between the anyon-driven superconductor and the ones described in the previous subsection? As argued in Ref. \cite{Shi2025_anyon_delocalization}, the effective energy functional (at $T = 0$)  for the disordered anyon-induced superconductor has the structure 
\begin{equation} 
H_{\rm avg} = \int d^2 x \kappa (\boldsymbol {\nabla} b_0)^2 - j_0(x) b_0 .
\end{equation} 
Here $b_0$ can be thought of as the vortex potential and $j_0(x)$ is a random `charge' distribution which arises from the presence of randomly localized anyons on top of which the mobile anyons form the superconductor. This also maps to the random XY model of Eqn. \ref{randomxy} but with {\it random flux}  rather than random phase shift. In other words, it is $\boldsymbol \nabla \times \boldsymbol a$ that is short-range correlated rather than $\boldsymbol a$ itself. The effective vortex potential in the random-flux XY model has very long-range correlations, and this makes it unstable toward forming a vortex glass at arbitrarily weak disorder, unlike the random phase shift model.

\section{Conclusion}

In this paper we have studied two distinct effects of disorder on chiral superconductors with non-trivial normal-state band geometry, motivated by the recent observation of chiral superconductivity in rhombohedral graphene and twisted MoTe$_2$ \cite{graphene_sc, Xu2025_SCdopeTMD}. Specifically, tMoTe$_2$ is more disordered than rhombohedral graphene, and has highly resistive normal states surrounding the superconductor, thereby making it important to consider disorder effects. 

Our first result concerns the mean-field suppression of $T_c$ by disorder. We showed that the standard Anderson theorem and Abrikosov-Gorkov pair-breaking formula are modified by band geometry in a way that can be captured by two dimensionless form factors $f_{\rm sym}$ and $f_m$, defined in Eqn.~\eqref{fsym and fm definition} as angular averages of the Bloch-state overlap $\Lambda_{\boldsymbol{k},\boldsymbol{p}}$. These form factors encode all the dependence on Berry curvature and quantum metric: in the limit of trivial band geometry $\Lambda_{\boldsymbol{k},\boldsymbol{p}} \to 1$, one recovers $f_{\rm sym} = 1$ and $f_m = \delta_{m,0}$, reproducing Anderson's theorem \cite{Anderson1959} for $m=0$ and the Abrikosov-Gorkov formula \cite{Mineev1999} for $m \neq 0$. For non-trivial geometry the effective pair-breaking strength is controlled by the combination $f_{\rm sym} - f_m$, and strong Berry curvature can therefore either enhance or suppress the robustness of a given pairing channel relative to the flat-band expectation. We showed that this disorder sensitivity can be expressed directly in terms of the residual resistivity $\rho$ through Eqn.~\eqref{Tc over rho}, making the result accessible to experiment without knowledge of the microscopic scattering time.

Applying this framework to the hole-doped tMoTe$_2$ Fermi surface near $\nu_h \approx 0.72$ using a realistic skyrmion crystal band structure \cite{Wu2019}, we find that the $m = \pm 1$ pairing channels are the most robust to disorder, with $f_m/f_{\rm sym} \approx 0.6$. The critical resistivities above which superconductivity is destroyed are of order several $\mathrm{k}\Omega$, indicating that a BCS instability of the large hole Fermi surface would require a relatively clean sample. The large measured residual resistivity in Ref.~\cite{Xu2025_SCdopeTMD} is thus in tension with the BCS interpretation of the observed superconductivity.

Our second result concerns the possibility of spontaneous vortex nucleation by disorder. For chiral superconductors, time-reversal breaking generates an orbital magnetization that couples linearly to a random scalar potential, leading to a tendency to induce random vorticity in the phase of the order parameter. However, we showed that this tendency is unable to destroy superconductivity for weak disorder, since this mechanism maps the long-wavelength physics onto the two-dimensional random-phase-shift $XY$ model of Eqn.~\eqref{randomxy}, whose long-range order is stable at weak disorder and which supports a sharp finite-temperature BKT transition \cite{Korshunov_2006}. This is in sharp contrast to superconductors stabilized by an anyonic fluid, where the appropriate model is the random-flux $XY$ model with long-range correlated disorder - a model that is unstable to arbitrarily weak randomness and supports only an Anomalous Vortex Glass phase with no finite-$T$ transition \cite{Shi2025_anyon_delocalization, Fisher_vortex-glass_1989, Fisher_thermal_1991}. Our result implies that if the chiral superconductivity in tMoTe$_2$ is of BCS origin, a weak random potential does not destroy the $T = 0$ long-range order or the associated finite-temperature phase transition. Observing signatures of a vortex glass phase - such as the absence of a sharp resistive transition - would therefore constitute evidence against a BCS mechanism and in favor of an anyonic origin.

Several directions remain open. The calculation of $T_c$ suppression here was carried out at mean-field level and neglects phase fluctuations, which may be important in two-dimensional systems near the superconductor-insulator transition. The interplay of the Berry-curvature-enhanced pair-breaking studied here with the geometric contribution to superfluid stiffness is an interesting question for future work. Additionally, in the RNG system the Fermi surface geometry and residual resistivity differ substantially from tMoTe$_2$, and a parallel analysis there - combined with the cleaner samples available - may provide a more direct test of the modified Abrikosov-Gorkov formula derived in Sec.~II.

We thank Yuan Cao, Pablo Jarillo-Herrero, Long Ju, Jeong Min Park, and Tingxin Li for useful discussions.  TS was supported by the U.S. Department of Energy under Grant DE-SC0008739. ISV acknowledges support by the Deutsche Forschungsgemeinschaft (DFG) through Research Grant Projects No. 542614019, No. 518372354, and No. 555335098. OpenAI's ChatGPT (GPT-6) was used to assist with language editing, literature searches, and verification of selected calculations. All AI-assisted suggestions were critically evaluated, and the authors independently verified the scientific results and references. The authors take full responsibility for the content of this manuscript.

\appendix

\section{A brief survey of candidate unconventional superconductors in $2d$ materials }\label{app: overview}

\begin{table*}[t]
\centering
\small
\setlength{\tabcolsep}{5pt}
\begin{tabular}{lcccccccccc}
\specialrule{1.5pt}{0pt}{0pt}
System & $\theta$ & $T_c$ & $\rho_0$ & $g$
& $k_F\ell$ & $\ell$ (nm) & $\xi$ (nm)
& $\ell/\xi$ & $\rho_0/R_0$ & Refs. \\
\specialrule{1.5pt}{0pt}{0pt}

tMoTe$_2$
& $3.83^\circ$ & $1.2$ K
& 5.0 k$\Omega$
& 1 & 10.3
& 14 & 23
& 0.62 & 0.19
& \cite{Xu2025_SCdopeTMD} \\

MATBG M1
& $1.16^\circ$ & $0.5$ K
& $6$ k$\Omega$
& 2 & 4.3
& 44
& 52
& 0.85 & 0.23
& \cite{Cao_unconventional_2018} \\

% tWSe$_2$ (it's VH singularity)
% & $5.0^\circ$ & $426$ mK
% & $150$ $\Omega$
% & 2 & 172 & 229 & 57 & 4.0 & 0.0058 & \cite{Guo_angle_2026}\\

tWSe$_2$ 
& $3.5^\circ$ & $120$ mK
& 0.5 $k\Omega$
& 2 & 51.6
& 147 & 69 & 2.1 & 0.02 & \cite{Xia_extended_2026} \\

tWSe$_2$ 
& $3.65^\circ$ & $180$ mK
& 0.5 $k\Omega$
& 2 & 51.6
& 141 & 52 & 2.7 & 0.02 & \cite{Xia_unconventional_2024} \\

MATTG ($\nu=-2-\delta$)
& $1.57^\circ$ & $2.9$ K
& 3.6 k$\Omega$
& 2 & 7.2
& 38
& 9.4 & 4.0
& 0.14
& \cite{Park_tunable_2021} \\

R4G (SC1)
& --- & $0.3$ K
& 0.8 k$\Omega$
& 1 & 65
& 259
& 21 & 12.3
& 0.03
& \cite{graphene_sc} \\

% tWSe$_2$(it's VH singularity)
% & $4.6^\circ$ & \textcolor{red}{---}
% & $<100$ $\Omega$
% & 2 & \textcolor{red}{258}
% & \textcolor{red}{398}
% & 34 & 11.7 & $<0.0039$ & \cite{Xia_simulating_2025} \\

\specialrule{1.5pt}{0pt}{0pt}
\end{tabular}

\caption{Normal-state transport parameters of selected
two-dimensional superconductors.
Here $\theta$ is the moiré twist angle, $T_c$ is superconducting critical temperature, $\rho_0$ is the residual resistivity, $g$ is the degeneracy of the electronic system, $\ell$ is the electronic mean free path due to impurity scattering, $\xi$ is coherence length and $R_0=h/e^2\simeq25.8\,\mathrm{k}\Omega$ is the resistance quantum. For tMoTe$_2$ we have assumed a conventional Drude metal with a Fermi surface determined by the deviation from $\nu = 1$.}
\label{tc for superconductors}
\end{table*}

Here we survey the superconducting critical temperatures, residual resistivities, and coherence lengths of prominent $2d$ materials that may be unconventional superconductors \cite{Andrei_the_2021, Andrei_graphene_2020}. Our primary goal is to compare their mean free paths $l$ with their superconducting coherence lengths $\xi$. These data are summarized in Table \ref{tc for superconductors}. We have not included non-moir\'e $2d$ TMD superconductors as they are generally in the clean limit. The residual resistivity is defined as the normal-state resistivity extrapolated to zero temperature, where it is primarily determined by disorder scattering. In this table it is either extracted from the $T = 0$ extrapolation of the $T$-dependent longitudinal resistivity $\rho_{xx} (T)$ or from $\rho_{xx} (B)$ at small temperatures when superconductivity is destroyed by magnetic field. Although these two methods may provide different residual resistivities $\rho_0$, the results are usually very close. The coherence length $\xi$ is extracted from the Ginzburg-Landau relation for the magnetic-field dependence of the critical temperature: $T_c/T_{c0}=1-2\pi B_{\perp} \xi^2/\Phi_0$, where $\Phi_0=h/(2e)$ is the superconducting flux quantum and $T_{c0}$ is the critical temperature at zero magnetic field. From the residual resistivity $\rho_0$ (and assuming a {\it standard} normal metal with Drude transport),  we obtain $k_F\ell=2 R_0/(g \rho_0)$ where $g$ is the flavor degeneracy of the quasiparticles. Finally using the approximate estimate of the Fermi momentum $k_F\approx \sqrt{4\pi n_{el}/g}$, we obtain an estimate of the electronic mean free path $\ell$.
\par 

All the candidate unconventional superconductors considered here have residual resistances smaller than the resistance quantum $R_0\approx25.8k\Omega$. But more importantly, all the superconductors in Table \ref{tc for superconductors} except for tMoTe$_2$ and magic angle twisted bilayer graphene (MATBG) are in the clean limit, meaning that the mean free path $\ell$ is larger than coherence length: $\ell>\xi$. In particular RnG (which is chiral) and twisted WSe$_2$ (whose pairing symmetry is not clear at present) both have large $l/\xi$. We have not included superconductivity in larger twist angle \cite{guo2025superconductivity,Guo_angle_2026} WSe$_2$ as it seems to be connected to a van Hove singularity which complicates extracting the mean free path from the normal state residual resistivity. 

The MATBG family of superconductors includes mutlilayer alternately twisted magic angle graphene systems. It is believed that the superconductor seen on MATTG, MAT4G, MAT5G are all essentially similar, and have the same pairing symmetry. However the higher layer members generally have lower resistivities and comparable, or higher T$_c$, than MATBG. Thus we see that MATTG is comfortably in the clean limit (as also are \cite{Park_robust_2022}  MAT4G and MAT5G which are not included in Table~\ref{tc for superconductors}). Superconductivity in MATBG itself survives up to the dirty limit, as seen from Table~\ref{tc for superconductors}. Indeed, MATBG superconductivity is also reported \cite{Cao_unconventional_2018} in samples with both higher $\rho_0$ and higher $T_c$ than those listed in Table~\ref{tc for superconductors}, suggesting a strong dependence on sample-specific properties.  Analysing the robustness of superconductivity to disorder in MATBG  requires knowledge of the band properties of the normal state just above $T_c$, which are substantially different from the bare band properties. This is, at the very least, evidenced by the the normal state Fermi surface having an area \cite{Cao_unconventional_2018} that corresponds to the doping away from filling $\nu = -2$, as opposed to the band filling. Thus a proper analysis will first have to incorporate the interaction effects responsible for the normal state band structure, the disorder coupling, and then study how the Anderson/Abrikosov-Gorkov theory is affected. This is beyond the scope of the present paper. 

For tMoTe$_2$, we used a carrier density determined by the deviation from $\nu = 1$, as is appropriate in a BCS mechanism.  Our analysis shows that the high residual resistivity creates a tension for a BCS mechanism arising from pairing quasiparticles of the band theory Fermi liquid, and allows us to argue that other mechanisms for unconventional superconductivity that are more robust to disorder (such as anyon-induced superconductivity \cite{Shi2024_doping,Shi2025_anyon_delocalization,Nosov2025_plateau,shi2026superconductivity,han2026single}) may be more likely. As a specific example, if the normal metallic state above $T_c$ is the ``singlet" $Z_3$ Orthogonal Metal \cite{senthil2026fractionalized,han2026single} born out of an anyon fluid, then there are two changes compared to the standard BCS treatment. First the mobile carrier density is determined by the deviation from $2/3$ and second the carriers are charge $1/3$ fermionic quasiparticles. The extracted mean free path of the charge $1/3$ fermionic quasiparticles is then about 33 times larger than that quoted in Table \ref{tc for superconductors}, while $\xi$ (as it is extracted from the field dependence of $T_c$) does not change which then  makes it strongly in the clean limit.

\bibliographystyle{unsrt}
\nocite{Wang_repulsion-driven_2026,lotrivc2026fractional}
\bibliography{references}

@article{Vojta2005,
  author  = {Florens, Serge and Vojta, Matthias},
  title   = {Impact of disorder on unconventional superconductors with competing ground states},
  journal = {Physical Review B},
  volume  = {71},
  pages   = {094516},
  year    = {2005},
}

@article{Anderson1959,
  author  = {Anderson, P. W.},
  title   = {Theory of Dirty Superconductors},
  journal = {Journal of Physics and Chemistry of Solids},
  volume  = {11},
  pages   = {26--30},
  year    = {1959}
}

@book{Mineev1999,
  author    = {Mineev, V. P. and Samokhin, K. V.},
  title     = {Introduction to Unconventional Superconductivity},
  publisher = {Gordon and Breach},
  address   = {Amsterdam},
  year      = {1999}
}

@article{Wu2019,
  title={Topological insulators in twisted transition metal dichalcogenide homobilayers},
  author={Wu, Fengcheng and Lovorn, Timothy and Tutuc, Emanuel and Martin, Ivar and MacDonald, Allan H},
  journal={Physical review letters},
  volume={122},
  number={8},
  pages={086402},
  year={2019},
  publisher={APS}
}

@article{Guerci_from_2025,
 author = {Guerci, Daniele and Abouelkomsan, Ahmed and Fu, Liang},
 issn = {1079-7114},
 journal = {Physical Review Letters},
 month = {October},
 number = {18},
 pages = {186601},
 publisher = {American Physical Society (APS)},
 title = {{From} {Fractionalization} to {Chiral} {Topological} {Superconductivity} in a {Flat} {Chern} {Band}},
 url = {http://dx.doi.org/10.1103/zm39-dstj},
 volume = {135},
 year = {2025}
}

@article{Fisher_vortex-glass_1989,
 author = {Fisher, Matthew P. A.},
 issn = {0031-9007},
 journal = {Physical Review Letters},
 month = {March},
 number = {12},
 pages = {1415--1418},
 publisher = {American Physical Society (APS)},
 title = {{Vortex-glass} superconductivity: {A} possible new phase in bulk high- ${\mathrm{T}}_{\mathrm{c}}$ oxides},
 url = {http://dx.doi.org/10.1103/PhysRevLett.62.1415},
 volume = {62},
 year = {1989}
}

@article{Korshunov_2006,
  title={Phase transitions in two-dimensional systems with continuous degeneracy},
  author={Korshunov, Sergei E},
  journal={Physics-Uspekhi},
  volume={49},
  number={3},
  pages={225--262},
  year={2006}
}

@article{Rubinstein_two-dimensionalmmlmath_1983,
 author = {Rubinstein, Michael and Shraiman, Boris and Nelson, David R.},
 issn = {0163-1829},
 journal = {Physical Review B},
 month = {February},
 number = {3},
 pages = {1800--1811},
 publisher = {American Physical Society (APS)},
 title = {{Two-dimensional} $\mathrm{XY}$ magnets with random {Dzyaloshinskii-Moriya} interactions},
 url = {http://dx.doi.org/10.1103/PhysRevB.27.1800},
 volume = {27},
 year = {1983}
}

@article{nattermann1995absence,
  title={Absence of reentrance in the two-dimensional XY-model with random phase shift},
  author={Nattermann, Thomas and Scheidl, Stefan and Korshunov, Sergey E and Li, Mai Suan},
  journal={Journal de Physique I},
  volume={5},
  number={5},
  pages={565--572},
  year={1995}
}

@article{scheidl1997glassy,
  title={Glassy vortex state in a two-dimensional disordered XY model},
  author={Scheidl, Stefan},
  journal={Physical Review B},
  volume={55},
  number={1},
  pages={457},
  year={1997},
  publisher={APS}
}

@article{carpentier1998disordered,
  title={Disordered XY models and Coulomb gases: renormalization via traveling waves},
  author={Carpentier, David and Le Doussal, Pierre},
  journal={Physical review letters},
  volume={81},
  number={12},
  pages={2558},
  year={1998},
  publisher={APS}
}

@article{Cao_unconventional_2018,
 author = {Cao, Yuan and Fatemi, Valla and Fang, Shiang and Watanabe, Kenji and Taniguchi, Takashi and Kaxiras, Efthimios and Jarillo-Herrero, Pablo},
 issn = {1476-4687},
 journal = {Nature},
 month = {March},
 number = {7699},
 pages = {43--50},
 publisher = {Springer Science and Business Media LLC},
 title = {{Unconventional} superconductivity in magic-angle graphene superlattices},
 url = {http://dx.doi.org/10.1038/nature26160},
 volume = {556},
 year = {2018}
}

@article{Andrei_the_2021,
 author = {Andrei, Eva Y. and Efetov, Dmitri K. and Jarillo-Herrero, Pablo and MacDonald, Allan H. and Mak, Kin Fai and Senthil, T. and Tutuc, Emanuel and Yazdani, Ali and Young, Andrea F.},
 issn = {2058-8437},
 journal = {Nature Reviews Materials},
 month = {March},
 number = {3},
 pages = {201--206},
 publisher = {Springer Science and Business Media LLC},
 title = {{The} marvels of moir\'{e} materials},
 url = {http://dx.doi.org/10.1038/s41578-021-00284-1},
 volume = {6},
 year = {2021}
}

@article{Andrei_graphene_2020,
 author = {Andrei, Eva Y. and MacDonald, Allan H.},
 issn = {1476-4660},
 journal = {Nature Materials},
 month = {November},
 number = {12},
 pages = {1265--1275},
 publisher = {Springer Science and Business Media LLC},
 title = {{Graphene} bilayers with a twist},
 url = {http://dx.doi.org/10.1038/s41563-020-00840-0},
 volume = {19},
 year = {2020}
}

@article{Sedov_quantum_2025,
  title={Quantum geometry and impurity sensitivity of superconductors without time-reversal symmetry: application to rhombohedral graphene and altermagnets},
  author={Sedov, Denis and Scheurer, Mathias S},
  journal={arXiv preprint arXiv:2510.19943},
  year={2025}
}

@article{Park_tunable_2021,
 author = {Park, Jeong Min and Cao, Yuan and Watanabe, Kenji and Taniguchi, Takashi and Jarillo-Herrero, Pablo},
 issn = {1476-4687},
 journal = {Nature},
 month = {February},
 number = {7845},
 pages = {249--255},
 publisher = {Springer Science and Business Media LLC},
 title = {{Tunable} strongly coupled superconductivity in magic-angle twisted trilayer graphene},
 url = {http://dx.doi.org/10.1038/s41586-021-03192-0},
 volume = {590},
 year = {2021}
}

@article{Xia_unconventional_2024,
  title={Unconventional superconductivity in twisted bilayer WSe$_2$ (2024)},
  author={Xia, Y and Han, Z and Watanabe, K and Taniguchi, T and Shan, J and Mak, KF},
  journal={arXiv preprint arXiv:2405.14784}
}

@article{graphene_sc,
 author = {Han, Tonghang and Lu, Zhengguang and Hadjri, Zach and Shi, Lihan and Wu, Zhenghan and Xu, Wei and Yao, Yuxuan and Cotten, Armel A. and Sharifi Sedeh, Omid and Weldeyesus, Henok and Yang, Jixiang and Seo, Junseok and Ye, Shenyong and Zhou, Muyang and Liu, Haoyang and Shi, Gang and Hua, Zhenqi and Watanabe, Kenji and Taniguchi, Takashi and Xiong, Peng and Zumb\"{u}hl, Dominik M. and Fu, Liang and Ju, Long},
 issn = {1476-4687},
 journal = {Nature},
 month = {May},
 number = {8072},
 pages = {654--661},
 publisher = {Springer Science and Business Media LLC},
 title = {{Signatures} of chiral superconductivity in rhombohedral graphene},
 url = {http://dx.doi.org/10.1038/s41586-025-09169-7},
 volume = {643},
 year = {2025}
}

@article{Fisher_thermal_1991,
 author = {Fisher, Daniel S. and Fisher, Matthew P. A. and Huse, David A.},
 issn = {1095-3795},
 journal = {Physical Review B},
 month = {January},
 number = {1},
 pages = {130--159},
 publisher = {American Physical Society (APS)},
 title = {{Thermal} fluctuations, quenched disorder, phase transitions, and transport in type-{II} superconductors},
 url = {http://dx.doi.org/10.1103/PhysRevB.43.130},
 volume = {43},
 year = {1991}
}

@ARTICLE{nandkishore2012orthogonal,
       author = {{Nandkishore}, Rahul and {Metlitski}, Max A. and {Senthil}, T.},
        title = "{Orthogonal metals: The simplest non-Fermi liquids}",
      journal = {\prb},
         year = 2012,
        month = jul,
       volume = {86},
       number = {4},
          eid = {045128},
        pages = {045128},
          doi = {10.1103/PhysRevB.86.045128},
archivePrefix = {arXiv},
       eprint = {1201.5998},
 primaryClass = {cond-mat.str-el},
       adsurl = {https://ui.adsabs.harvard.edu/abs/2012PhRvB..86d5128N}
}

@ARTICLE{Park2023_FQAH_TMD,
       author = {{Park}, Heonjoon and {Cai}, Jiaqi and {Anderson}, Eric and {Zhang}, Yinong and {Zhu}, Jiayi and {Liu}, Xiaoyu and {Wang}, Chong and {Holtzmann}, William and {Hu}, Chaowei and {Liu}, Zhaoyu and {Taniguchi}, Takashi and {Watanabe}, Kenji and {Chu}, Jiun-Haw and {Cao}, Ting and {Fu}, Liang and {Yao}, Wang and {Chang}, Cui-Zu and {Cobden}, David and {Xiao}, Di and {Xu}, Xiaodong},
        title = "{Observation of fractionally quantized anomalous Hall effect}",
      journal = {\nat},
         year = 2023,
        month = oct,
       volume = {622},
       number = {7981},
        pages = {74-79},
          doi = {10.1038/s41586-023-06536-0},
archivePrefix = {arXiv},
       eprint = {2308.02657},
 primaryClass = {cond-mat.mes-hall},
       adsurl = {https://ui.adsabs.harvard.edu/abs/2023Natur.622...74P}
}

@ARTICLE{Cai2023_FQAHTMD,
       author = {{Cai}, Jiaqi and {Anderson}, Eric and {Wang}, Chong and {Zhang}, Xiaowei and {Liu}, Xiaoyu and {Holtzmann}, William and {Zhang}, Yinong and {Fan}, Fengren and {Taniguchi}, Takashi and {Watanabe}, Kenji and {Ran}, Ying and {Cao}, Ting and {Fu}, Liang and {Xiao}, Di and {Yao}, Wang and {Xu}, Xiaodong},
        title = "{Signatures of fractional quantum anomalous Hall states in twisted MoTe$_{2}$}",
      journal = {\nat},
         year = 2023,
        month = oct,
       volume = {622},
       number = {7981},
        pages = {63-68},
          doi = {10.1038/s41586-023-06289-w},
archivePrefix = {arXiv},
       eprint = {2304.08470},
 primaryClass = {cond-mat.mes-hall},
       adsurl = {https://ui.adsabs.harvard.edu/abs/2023Natur.622...63C}
}

@ARTICLE{Xu2023_FQAHTMD,
       author = {{Xu}, Fan and {Sun}, Zheng and {Jia}, Tongtong and {Liu}, Chang and {Xu}, Cheng and {Li}, Chushan and {Gu}, Yu and {Watanabe}, Kenji and {Taniguchi}, Takashi and {Tong}, Bingbing and {Jia}, Jinfeng and {Shi}, Zhiwen and {Jiang}, Shengwei and {Zhang}, Yang and {Liu}, Xiaoxue and {Li}, Tingxin},
        title = "{Observation of Integer and Fractional Quantum Anomalous Hall Effects in Twisted Bilayer MoTe$_{2}$}",
      journal = {Physical Review X},
         year = 2023,
        month = jul,
       volume = {13},
       number = {3},
          eid = {031037},
        pages = {031037},
          doi = {10.1103/PhysRevX.13.031037},
archivePrefix = {arXiv},
       eprint = {2308.06177},
 primaryClass = {cond-mat.mes-hall},
       adsurl = {https://ui.adsabs.harvard.edu/abs/2023PhRvX..13c1037X}
}

@ARTICLE{Lu2023_FQAHPenta,
       author = {{Lu}, Zhengguang and {Han}, Tonghang and {Yao}, Yuxuan and {Reddy}, Aidan P. and {Yang}, Jixiang and {Seo}, Junseok and {Watanabe}, Kenji and {Taniguchi}, Takashi and {Fu}, Liang and {Ju}, Long},
        title = "{Fractional quantum anomalous Hall effect in multilayer graphene}",
      journal = {\nat},
         year = 2024,
        month = feb,
       volume = {626},
       number = {8000},
        pages = {759-764},
          doi = {10.1038/s41586-023-07010-7},
archivePrefix = {arXiv},
       eprint = {2309.17436},
 primaryClass = {cond-mat.mes-hall},
       adsurl = {https://ui.adsabs.harvard.edu/abs/2024Natur.626..759L}
}

@ARTICLE{Lu2025_EQAH,
       author = {{Lu}, Zhengguang and {Han}, Tonghang and {Yao}, Yuxuan and {Hadjri}, Zach and {Yang}, Jixiang and {Seo}, Junseok and {Shi}, Lihan and {Ye}, Shenyong and {Watanabe}, Kenji and {Taniguchi}, Takashi and {Ju}, Long},
        title = "{Extended quantum anomalous Hall states in graphene/hBN moir{\'e} superlattices}",
      journal = {\nat},
         year = 2025,
        month = jan,
       volume = {637},
       number = {8048},
        pages = {1090-1095},
          doi = {10.1038/s41586-024-08470-1},
archivePrefix = {arXiv},
       eprint = {2408.10203},
 primaryClass = {cond-mat.mes-hall},
       adsurl = {https://ui.adsabs.harvard.edu/abs/2025Natur.637.1090L}
}

@article{Zeng2023_FQAHTMD,
	author = {Zeng, Yihang and Xia, Zhengchao and Kang, Kaifei and Zhu, Jiacheng and Kn{\"u}ppel, Patrick and Vaswani, Chirag and Watanabe, Kenji and Taniguchi, Takashi and Mak, Kin Fai and Shan, Jie},
	da = {2023/10/01},
	doi = {10.1038/s41586-023-06452-3},
	id = {Zeng2023},
	isbn = {1476-4687},
	journal = {Nature},
	number = {7981},
	pages = {69--73},
	title = {Thermodynamic evidence of fractional Chern insulator in moir\'e MoTe2},
	ty = {JOUR},
	url = {https://doi.org/10.1038/s41586-023-06452-3},
	volume = {622},
	year = {2023}}

@ARTICLE{Xu2025_SCdopeTMD,
       author = {{Xu}, Fan and {Sun}, Zheng and {Li}, Jiayi and {Zheng}, Ce and {Xu}, Cheng and {Gao}, Jingjing and {Jia}, Tongtong and {Watanabe}, Kenji and {Taniguchi}, Takashi and {Tong}, Bingbing and {Lu}, Li and {Jia}, Jinfeng and {Shi}, Zhiwen and {Jiang}, Shengwei and {Zhang}, Yuanbo and {Zhang}, Yang and {Lei}, Shiming and {Liu}, Xiaoxue and {Li}, Tingxin},
        title = "{Signatures of unconventional superconductivity near reentrant and fractional quantum anomalous Hall insulators}",
      journal = {arXiv e-prints},
         year = 2025,
        month = apr,
          eid = {arXiv:2504.06972},
        pages = {arXiv:2504.06972},
          doi = {10.48550/arXiv.2504.06972},
archivePrefix = {arXiv},
       eprint = {2504.06972},
 primaryClass = {cond-mat.mes-hall},
       adsurl = {https://ui.adsabs.harvard.edu/abs/2025arXiv250406972X}
}

@article{Laughlin1988_anyonSC,
  title = {Superconducting Ground State of Noninteracting Particles Obeying Fractional Statistics},
  author = {Laughlin, R. B.},
  journal = {Phys. Rev. Lett.},
  volume = {60},
  issue = {25},
  pages = {2677--2680},
  numpages = {0},
  year = {1988},
  month = {Jun},
  publisher = {American Physical Society},
  doi = {10.1103/PhysRevLett.60.2677},
  url = {https://link.aps.org/doi/10.1103/PhysRevLett.60.2677}
}

@article{Fetter1989_anyonSC_RPA,
  title = {Random-phase approximation in the fractional-statistics gas},
  author = {Fetter, A. L. and Hanna, C. B. and Laughlin, R. B.},
  journal = {Phys. Rev. B},
  volume = {39},
  issue = {13},
  pages = {9679--9681},
  numpages = {0},
  year = {1989},
  month = {May},
  publisher = {American Physical Society},
  doi = {10.1103/PhysRevB.39.9679},
  url = {https://link.aps.org/doi/10.1103/PhysRevB.39.9679}
}

@article{Lee1989_anyonSC,
  title = {Anyon superconductivity and the fractional quantum Hall effect},
  author = {Lee, Dung-Hai and Fisher, Matthew P. A.},
  journal = {Phys. Rev. Lett.},
  volume = {63},
  issue = {8},
  pages = {903--906},
  numpages = {0},
  year = {1989},
  month = {Aug},
  publisher = {American Physical Society},
  doi = {10.1103/PhysRevLett.63.903},
  url = {https://link.aps.org/doi/10.1103/PhysRevLett.63.903}
}

@ARTICLE{Chen1989_anyonSC,
       author = {{Chen}, Yi-Hong and {Wilczek}, Frank and {Witten}, Edward and {Halperin}, Bertrand I.},
        title = "{On Anyon Superconductivity}",
      journal = {International Journal of Modern Physics A},
         year = 1989,
        month = jan,
       volume = {4},
       number = {15},
        pages = {3983},
          doi = {10.1142/S0217751X89001631},
       adsurl = {https://ui.adsabs.harvard.edu/abs/1989IJMPA...4.3983C}
}

@article{Wen1990_anyonSC,
  title = {Compressibility and superfluidity in the fractional-statistics liquid},
  author = {Wen, X. G. and Zee, A.},
  journal = {Phys. Rev. B},
  volume = {41},
  issue = {1},
  pages = {240--253},
  numpages = {0},
  year = {1990},
  month = {Jan},
  publisher = {American Physical Society},
  doi = {10.1103/PhysRevB.41.240},
  url = {https://link.aps.org/doi/10.1103/PhysRevB.41.240}
}

@article{Kalantre_fermiology_2026,
  title={Fermiology and the candidate chiral superconductor in rhombohedral tetralayer graphene},
  author={Kalantre, Sandesh S and Alexander, Ben H and May-Mann, Julian and Herzog-Arbeitman, Jonah and Hocking, Marisa and Cao, Qingrui and Watanabe, Kenji and Taniguchi, Takashi and Goldhaber-Gordon, David and Mannix, Andrew J and others},
  journal={arXiv preprint arXiv:2606.05356},
  year={2026}
}

@article{Roy_band_2014,
 author = {Roy, Rahul},
 issn = {1550-235X},
 journal = {Physical Review B},
 month = {October},
 number = {16},
 pages = {165139},
 publisher = {American Physical Society (APS)},
 title = {{Band} geometry of fractional topological insulators},
 url = {http://dx.doi.org/10.1103/PhysRevB.90.165139},
 volume = {90},
 year = {2014}
}

@article{Oh_evidence_2021,
 author = {Oh, Myungchul and Nuckolls, Kevin P. and Wong, Dillon and Lee, Ryan L. and Liu, Xiaomeng and Watanabe, Kenji and Taniguchi, Takashi and Yazdani, Ali},
 issn = {1476-4687},
 journal = {Nature},
 month = {October},
 number = {7888},
 pages = {240--245},
 publisher = {Springer Science and Business Media LLC},
 title = {{Evidence} for unconventional superconductivity in twisted bilayer graphene},
 url = {http://dx.doi.org/10.1038/s41586-021-04121-x},
 volume = {600},
 year = {2021}
}

@article{Kim_evidence_2022,
 author = {Kim, Hyunjin and Choi, Youngjoon and Lewandowski, Cyprian and Thomson, Alex and Zhang, Yiran and Polski, Robert and Watanabe, Kenji and Taniguchi, Takashi and Alicea, Jason and Nadj-Perge, Stevan},
 issn = {1476-4687},
 journal = {Nature},
 month = {June},
 number = {7914},
 pages = {494--500},
 publisher = {Springer Science and Business Media LLC},
 title = {{Evidence} for unconventional superconductivity in twisted trilayer graphene},
 url = {http://dx.doi.org/10.1038/s41586-022-04715-z},
 volume = {606},
 year = {2022}
}

@article{Park_robust_2022,
 author = {Park, Jeong Min and Cao, Yuan and Xia, Li-Qiao and Sun, Shuwen and Watanabe, Kenji and Taniguchi, Takashi and Jarillo-Herrero, Pablo},
 issn = {1476-4660},
 journal = {Nature Materials},
 month = {July},
 number = {8},
 pages = {877--883},
 publisher = {Springer Science and Business Media LLC},
 title = {{Robust} superconductivity in magic-angle multilayer graphene family},
 url = {http://dx.doi.org/10.1038/s41563-022-01287-1},
 volume = {21},
 year = {2022}
}

@article{Lake_pairing_2022,
 author = {Lake, Ethan and Patri, Adarsh S. and Senthil, T.},
 issn = {2469-9969},
 journal = {Physical Review B},
 month = {September},
 number = {10},
 pages = {104506},
 publisher = {American Physical Society (APS)},
 title = {{Pairing} symmetry of twisted bilayer graphene: {A} phenomenological synthesis},
 url = {http://dx.doi.org/10.1103/PhysRevB.106.104506},
 volume = {106},
 year = {2022}
}

@article{Biswas_andreev_2025,
 author = {Biswas, Sayak and Suman, Saurav and Randeria, Mohit and Sensarma, Rajdeep},
 issn = {1091-6490},
 journal = {Proceedings of the National Academy of Sciences},
 month = {November},
 number = {46},
 pages = {e2509881122},
 publisher = {National Academy of Sciences},
 title = {{Andreev} versus tunneling spectroscopy of unconventional flat-band superconductors},
 url = {http://dx.doi.org/10.1073/pnas.2509881122},
 volume = {122},
 year = {2025}
}

@article{Park_experimental_2026,
 author = {Park, Jeong Min and Sun, Shuwen and Watanabe, Kenji and Taniguchi, Takashi and Jarillo-Herrero, Pablo},
 issn = {1095-9203},
 journal = {Science},
 month = {January},
 number = {6780},
 pages = {79--83},
 publisher = {American Association for the Advancement of Science (AAAS)},
 title = {{Experimental} evidence for nodal superconducting gap in moir\'{e} graphene},
 url = {http://dx.doi.org/10.1126/science.adv8376},
 volume = {391},
 year = {2026}
}

@article{Sun_nodal_2026,
  title={Nodal nematic superconductivity in moir\'e graphene},
  author={Sun, Shuwen and Palle, Grgur and Watanabe, Kenji and Taniguchi, Takashi and Fernandes, Rafael M and Jarillo-Herrero, Pablo and Park, Jeong Min},
  journal={arXiv preprint arXiv:2610.20605},
  year={2026}
}

@article{Biswas_probing_2026,
  title={Probing the pairing symmetry of moir\'e graphene superconductors},
  author={Biswas, Sayak and Sensarma, Rajdeep and Randeria, Mohit},
  journal={arXiv preprint arXiv:2606.23994},
  year={2026}
}

@article{Hua_multi-knob_2026,
  title={Multi-Knob Switchable Chiral Superconductivity Quartet in Rhombohedral Graphene},
  author={Hua, Zhenqi and Ye, Shenyong and Pattanakanvijit, Phatthanon and Shi, Gang and Han, Tonghang and Aitken, Emily and Yang, Jixiang and Seo, Junseok and Liu, Haoyang and Hao, Ran and others},
  journal={arXiv preprint arXiv:2607.06520},
  year={2026}
}

@article{Xia_extended_2026,
  title={Extended strange metal and superconductivity in partially filled moir\'e Hubbard bands},
  author={Xia, Yiyu and Zhang, Yichi and Han, Zhongdong and Li, Hongyuan and Bauer, Brandon and Watanabe, Kenji and Taniguchi, Takashi and Zhao, Wenjin and Shan, Jie and Mak, Kin Fai},
  journal={arXiv preprint arXiv:2610.22307},
  year={2026}
}

@article{Guo_angle_2026,
 author = {Guo, Yinjie and Cenker, John and Fischer, Ammon and Mu\~{n}oz-Segovia, Daniel and Pack, Jordan and Holtzman, Luke and Klebl, Lennart and Watanabe, Kenji and Taniguchi, Takashi and Barmak, Katayun and Hone, James and Rubio, Angel and Kennes, Dante M. and Millis, Andrew J. and Pasupathy, Abhay and Dean, Cory R.},
 issn = {1476-4687},
 journal = {Nature},
 month = {April},
 number = {8110},
 pages = {622--627},
 publisher = {Springer Science and Business Media LLC},
 title = {{Angle} evolution of the superconducting phase diagram in twisted bilayer $\mathrm{W}\mathrm{Se}_{2}$},
 url = {http://dx.doi.org/10.1038/s41586-026-10357-2},
 volume = {652},
 year = {2026}
}

@article{guo2025superconductivity,
  title={Superconductivity in 5.0 twisted bilayer WSe2},
  author={Guo, Yinjie and Pack, Jordan and Swann, Joshua and Holtzman, Luke and Cothrine, Matthew and Watanabe, Kenji and Taniguchi, Takashi and Mandrus, David G and Barmak, Katayun and Hone, James and others},
  journal={Nature},
  volume={637},
  number={8047},
  pages={839--845},
  year={2025},
  publisher={Nature Publishing Group UK, London}
}

@ARTICLE{Tang2013_anyonSC,
       author = {{Tang}, Evelyn and {Wen}, Xiao-Gang},
        title = "{Superconductivity with intrinsic topological order induced by pure Coulomb interaction and time-reversal symmetry breaking}",
      journal = {\prb},
         year = 2013,
        month = nov,
       volume = {88},
       number = {19},
          eid = {195117},
        pages = {195117},
          doi = {10.1103/PhysRevB.88.195117},
archivePrefix = {arXiv},
       eprint = {1306.1528},
 primaryClass = {cond-mat.str-el},
       adsurl = {https://ui.adsabs.harvard.edu/abs/2013PhRvB..88s5117T}
}

@ARTICLE{Shi2024_doping,
       author = {{Shi}, Zhengyan Darius and {Senthil}, T.},
        title = "{Doping a Fractional Quantum Anomalous Hall Insulator}",
      journal = {Physical Review X},
         year = 2025,
        month = jul,
       volume = {15},
       number = {3},
          eid = {031069},
        pages = {031069},
          doi = {10.1103/kcm5-hx56},
archivePrefix = {arXiv},
       eprint = {2409.20567},
 primaryClass = {cond-mat.str-el},
       adsurl = {https://ui.adsabs.harvard.edu/abs/2025PhRvX..15c1069S}
}

@ARTICLE{Kim2024_anyonSC,
       author = {{Kim}, Minho and {Timmel}, Abigail and {Ju}, Long and {Wen}, Xiao-Gang},
        title = "{Topological chiral superconductivity beyond pairing in a Fermi liquid}",
      journal = {\prb},
         year = 2025,
        month = jan,
       volume = {111},
       number = {1},
          eid = {014508},
        pages = {014508},
          doi = {10.1103/PhysRevB.111.014508},
archivePrefix = {arXiv},
       eprint = {2409.18067},
 primaryClass = {cond-mat.str-el},
       adsurl = {https://ui.adsabs.harvard.edu/abs/2025PhRvB.111a4508K}
}

@ARTICLE{Shi2025_dopeMR,
       author = {{Darius Shi}, Zhengyan and {Zhang}, Carolyn and {Senthil}, T.},
        title = "{Doping lattice non-abelian quantum Hall states}",
      journal = {SciPost Physics},
       volume = {19},
         year = 2025,
        month = dec,
          eid = {150},
        pages = {150},
          doi = {10.21468/scipostphys.19.6.150},
archivePrefix = {arXiv},
       eprint = {2505.02893},
 primaryClass = {cond-mat.str-el},
       adsurl = {https://ui.adsabs.harvard.edu/abs/2025arXiv250502893D}
}

@ARTICLE{Shi2025_anyon_delocalization,
       author = {{Darius Shi}, Zhengyan and {Senthil}, T.},
        title = "{Anyon delocalization transitions out of a disordered fractional quantum anomalous Hall insulator}",
      journal = {Proceedings of the National Academy of Sciences},
       volume = {122},
         year = 2025,
        month = dec,
          eid = {e2520608122},
        pages = {e2520608122},
          doi = {10.1073/pnas.2520608122},
archivePrefix = {arXiv},
       eprint = {2506.02128},
 primaryClass = {cond-mat.str-el},
       adsurl = {https://ui.adsabs.harvard.edu/abs/2025arXiv250602128D}
}

@ARTICLE{Nosov2025_plateau,
       author = {{Nosov}, Pavel A. and {Han}, Zhaoyu and {Khalaf}, Eslam},
        title = "{Anyon superconductivity and plateau transitions in doped fractional quantum anomalous Hall insulators}",
      journal = {\prl},
       volume = {136},
         year = 2026,
        month = mar,
          eid = {106501},
        pages = {106501},
          doi = {10.1103/6bgj-bfdn},
archivePrefix = {arXiv},
       eprint = {2506.02108},
 primaryClass = {cond-mat.str-el},
       adsurl = {https://ui.adsabs.harvard.edu/abs/2025arXiv250602108N}
}

@ARTICLE{Pichler2025_anyonSC,
       author = {{Pichler}, Fabian and {Kuhlenkamp}, Clemens and {Knap}, Michael and {Vishwanath}, Ashvin},
        title = "{Microscopic Mechanism of Anyon Superconductivity Emerging from Fractional Chern Insulators}",
      journal = {Newton},
       volume = {2},
         year = 2026,
        month = mar,
          eid = {100340},
        pages = {100340},
          doi = {10.1016/j.newton.2025.100340},
archivePrefix = {arXiv},
       eprint = {2506.08000},
 primaryClass = {cond-mat.str-el},
       adsurl = {https://ui.adsabs.harvard.edu/abs/2025arXiv250608000P}
}

@ARTICLE{Han2025_anyonexciton,
       author = {{Han}, Zhaoyu and {Wang}, Taige and {Dong}, Zhihuan and {Zaletel}, Michael P. and {Vishwanath}, Ashvin},
        title = "{Anyon superfluidity of excitons in quantum Hall bilayers}",
      journal = {Physical Review X},
       volume = {16},
         year = 2026,
        month = may,
          eid = {021044},
        pages = {021044},
          doi = {10.1103/xs7g-dmn8},
archivePrefix = {arXiv},
       eprint = {2508.14894},
 primaryClass = {cond-mat.str-el},
       adsurl = {https://ui.adsabs.harvard.edu/abs/2025arXiv250814894H}
}

@ARTICLE{Kuhlenkamp2025_HofHubb,
       author = {{Kuhlenkamp}, Clemens and {Divic}, Stefan and {Zaletel}, Michael P. and {Soejima}, Tomohiro and {Vishwanath}, Ashvin},
        title = "{Robust superconductivity upon doping chiral spin liquid and Chern insulators in a Hubbard-Hofstadter model}",
      journal = {arXiv e-prints},
         year = 2025,
        month = sep,
          eid = {arXiv:2509.02675},
        pages = {arXiv:2509.02675},
          doi = {10.48550/arXiv.2509.02675},
archivePrefix = {arXiv},
       eprint = {2509.02675},
 primaryClass = {cond-mat.str-el},
       adsurl = {https://ui.adsabs.harvard.edu/abs/2025arXiv250902675K}
}

@ARTICLE{lotrivc2026phases,
       author = {Lotri\v{c}, Tev\v{z} and Simon, Steven H.},
        title = "{Phases of itinerant anyons in Laughlin's quantum Hall states on a lattice}",
      journal = {arXiv e-prints},
         year = 2026,
        month = mar,
          eid = {arXiv:2603.22389},
        pages = {arXiv:2603.22389},
          doi = {10.48550/arXiv.2603.22389},
archivePrefix = {arXiv},
       eprint = {2603.22389},
 primaryClass = {cond-mat.str-el},
       adsurl = {https://ui.adsabs.harvard.edu/abs/2026arXiv260322389L}
}

@ARTICLE{Shi2025_nonAbelian_TSC,
       author = {{Darius Shi}, Zhengyan and {Senthil}, T.},
        title = "{Non-Abelian topological superconductivity from melting Abelian fractional Chern insulators}",
      journal = {arXiv e-prints},
         year = 2025,
        month = dec,
          eid = {arXiv:2512.17996},
        pages = {arXiv:2512.17996},
          doi = {10.48550/arXiv.2512.17996},
archivePrefix = {arXiv},
       eprint = {2512.17996},
 primaryClass = {cond-mat.str-el},
       adsurl = {https://ui.adsabs.harvard.edu/abs/2025arXiv251217996D}
}

@ARTICLE{Fan2026_weakpairing_SC,
       author = {{Fan}, Zheng-Duo and {Vishwanath}, Ashvin and {Wang}, Zijian},
        title = "{Hidden weak-pairing superconductivity of non-interacting anyons obeying $\frac{1}{3}$ statistics}",
      journal = {arXiv e-prints},
         year = 2026,
        month = may,
          eid = {arXiv:2605.19036},
        pages = {arXiv:2605.19036},
          doi = {10.48550/arXiv.2605.19036},
archivePrefix = {arXiv},
       eprint = {2605.19036},
 primaryClass = {cond-mat.str-el},
       adsurl = {https://ui.adsabs.harvard.edu/abs/2026arXiv260519036F}
}

@ARTICLE{Wang2026_U3anyonSC,
       author = {{Wang}, Taige},
        title = "{Topological superconductivity from Abelian fractional Chern insulators}",
      journal = {arXiv e-prints},
         year = 2026,
        month = may,
          eid = {arXiv:2605.29034},
        pages = {arXiv:2605.29034},
          doi = {10.48550/arXiv.2605.29034},
archivePrefix = {arXiv},
       eprint = {2605.29034},
 primaryClass = {cond-mat.str-el},
       adsurl = {https://ui.adsabs.harvard.edu/abs/2026arXiv260529034W}
}

@article{shi2026superconductivity,
  title={Superconductivity and non-Fermi liquid metals in a charge-1/3 anyon fluid},
  author={Shi, Zhengyan Darius and Senthil, T},
  journal={arXiv preprint arXiv:2606.20403},
  year={2026}
}

@article{senthil2026fractionalized,
  title={Fractionalized metals from doped anyons: Application to tMoTe2},
  author={Senthil, T},
  journal={arXiv preprint arXiv:2607.11484},
  year={2026}
}

@article{han2026single,
  title={Single-component twisted Z3  orthogonal metal in an e/3-anyon fluid},
  author={Han, Zhaoyu and Vishwanath, Ashvin and Khalaf, Eslam},
  journal={arXiv preprint arXiv:2607.15274},
  year={2026}
}

@article{zhang2026color,
  title={Color superconductors and holon metals from doping a Fractional Chern insulator},
  author={Zhang, Ya-Hui},
  journal={arXiv preprint arXiv:2607.18238},
  year={2026}
}

@article{mehta2026coloring,
  title={Coloring in anyon superconductivity},
  author={Mehta, Umang and Nakajima, Yuto and Goldman, Hart},
  journal={arXiv preprint arXiv:2607.19470},
  year={2026}
}

@article{lotrivc2026fractional,
  title={Is the fractional Chern insulator-superconductor transition in twisted MoTe $ \_2 $ direct?},
  author={Lotri{\v{c}}, Tev{\v{z}} and Simon, Steven H},
  journal={arXiv preprint arXiv:2609.30368},
  year={2026}
}

@article{xu2025chiral,
  title={Chiral superconductivity from spin polarized Chern band in twisted MoTe 2},
  author={Xu, Cheng and Zou, Nianlong and Peshcherenko, Nikolai and Jahin, Ammar and Li, Tingxin and Lin, Shi-Zeng and Zhang, Yang},
  journal={Physical Review Letters},
  volume={135},
  number={26},
  pages={266005},
  year={2025},
  publisher={APS}
}

@article{Wang_repulsion-driven_2026,
    author = {Wang, T. and Soejima, T. and Yao, N. Y. and Vishwanath, A. and Zaletel, M. P. and Wang, T. and Parker, D. E.},
    title = {Repulsion-Driven Topological Superconductivity in Twisted {MoTe}$_2$},
    journal = {arXiv preprint arXiv:2611.03833},
    year = {2026},
    eprint = {2611.03833},
    archivePrefix = {arXiv}
}

@article{guerci2026topological,
  title={Topological superconductivity with emergent vortex lattice in twisted semiconductors},
  author={Guerci, Daniele and Abouelkomsan, Ahmed and Fu, Liang},
  journal={arXiv preprint arXiv:2602.15106},
  year={2026}
}

\end{document}